\documentclass[twocolumn,floatfix, nofootinbib,prd,aps,preprintnumbers,showpacs,showkeys,superscriptaddress,longbibliography]{revtex4-1}

\usepackage[mathscr]{euscript}
\usepackage{amsmath}
\usepackage{graphicx}
\usepackage{dcolumn}
\usepackage{bm}
\usepackage{epsfig}
\usepackage{amssymb,latexsym,mathrsfs}
\usepackage{graphicx}
\usepackage{color}
\usepackage{hyperref}
\usepackage{float}
\usepackage{diagbox}

\usepackage{tikz}

\newcommand{\xmark}{\ding{55}}
\usepackage{amssymb}
\usepackage{pifont}%

\usepackage{amsmath}
\usepackage{physics}
\usepackage[style=english]{csquotes} 

\usepackage{lipsum, babel}
\usepackage{amssymb}
\usepackage{color}
\usepackage{hyperref}
\usepackage{bm}
\usepackage{amsfonts}
\usepackage{mathrsfs}
\usepackage{graphicx}
\usepackage{amsmath}
\usepackage{float}
\usepackage{braket}
\usepackage{natbib}

     \hypersetup{
    colorlinks=true,
    linkcolor=red,
    citecolor=blue,
} 

\usepackage[caption=false]{subfig}

\usepackage{amssymb}
\usepackage[mathscr]{euscript}
\usepackage{amsmath}
\usepackage{graphicx}
\usepackage{dcolumn}
\usepackage{bm}
\usepackage{epsfig}
\usepackage{amssymb,latexsym,mathrsfs}
\usepackage{graphicx}
\usepackage{color}
\usepackage{hyperref}
\usepackage{float}
\usepackage[thinlines]{easytable}
\usepackage{multirow}
\usepackage{diagbox}
\usepackage{varwidth}
\usepackage{array}
\newcolumntype{P}[1]{>{\centering\arraybackslash}p{#1}}

\newcolumntype{M}[1]{>{\centering\arraybackslash}m{#1}}
\newcolumntype{N}{@{}m{0pt}@{}}

\usepackage{amsmath}

\usepackage{amssymb}
\usepackage{pifont}
\hypersetup{
    colorlinks=true,
    linkcolor=red,
    citecolor=blue,
}

\usepackage{amsmath}
\usepackage{amssymb}
\usepackage[caption=false]{subfig}
\usepackage{hyperref}
\usepackage{url}
\usepackage{xcolor}
\usepackage{color}
\definecolor{amaranth}{rgb}{0.9, 0.17, 0.31}
\definecolor{purple(munsell)}{rgb}{0.62, 0.0, 0.77}
\definecolor{americanrose}{rgb}{1.0, 0.01, 0.24}
\definecolor{palatinateblue}{rgb}{0.15, 0.23, 0.89}
\definecolor{royalblue(web)}{rgb}{0.25, 0.41, 0.88}
\definecolor{hanpurple}{rgb}{0.32, 0.09, 0.98}
\definecolor{beaublue}{rgb}{0.74, 0.83, 0.9}
\definecolor{carminered}{rgb}{1.0, 0.0, 0.22}
\definecolor{brightpink}{rgb}{1.0, 0.0, 0.5}
\definecolor{vividviolet}{rgb}{0.62, 0.0, 1.0}

\definecolor{electron}{rgb}{1.0, 0.67, 0.22}
\hypersetup{ linktoc=all,
    colorlinks, linkcolor={palatinateblue},
    citecolor={brightpink}, urlcolor={amaranth}}

\usepackage{physics}
\usepackage{csquotes}
\usepackage{mathtools}
\usepackage{soul}
\makeatletter
\def\l@subsubsection#1#2{}
\makeatother

\begin{document}

\newcommand*\diff{\mathop{}\!\mathrm{d}}
\newcommand*\Diff[1]{\mathop{}\!\mathrm{d^#1}}
\renewcommand{\d}[1]{\ensuremath{\operatorname{d}\!{#1}}}
\newcommand{\be}{\begin{equation}}
\newcommand{\ee}{\end{equation}}
\newcommand{\bs}{\begin{split}} 
\newcommand{\bea}{\begin{eqnarray}}
\newcommand{\eea}{\end{eqnarray}}
\newcommand{\mike}[1]{\textcolor{violet}{[{\bf MG}: #1]}} 
\newcommand{\morgan}[1]{\textcolor{purple}{[{\bf ML}: #1]}}
\newcommand{\ievlev}[1]{\textcolor{red}{[{\bf EI}: #1]}}

\newcommand{\sgn}{\operatorname{sgn}}
\newcommand{\hhat}[1]{\hat {\hat{#1}}}
\newcommand{\pslash}[1]{#1\llap{\sl/}}
\newcommand{\kslash}[1]{\rlap{\sl/}#1}
\newcommand{\lab}[1]{}
\newcommand{\iref}[2]{}
\newcommand{\sto}[1]{\begin{center} \textit{#1} \end{center}}
\newcommand{\rf}[1]{{\color{blue}[\textit{#1}]}}
\newcommand{\eml}[1]{#1}
\newcommand{\el}[1]{\label{#1}}
\newcommand{\er}[1]{Eq.\eqref{#1}}
\newcommand{\df}[1]{\textbf{#1}}
\newcommand{\mdf}[1]{\pmb{#1}}
\newcommand{\ft}[1]{\footnote{#1}}
\newcommand{\n}[1]{$#1$}
\newcommand{\fals}[1]{$^\times$ #1}
\newcommand{\new}{{\color{red}$^{NEW}$ }}
\newcommand{\ci}[1]{}
\newcommand{\de}[1]{{\color{green}\underline{#1}}}
\newcommand{\ke}{\rangle}
\newcommand{\br}{\langle}
\newcommand{\lb}{\left(}
\newcommand{\rb}{\right)}
\newcommand{\lbk}{\left[}
\newcommand{\rbk}{\right]}
\newcommand{\blb}{\Big(}
\newcommand{\brb}{\Big)}
\newcommand{\nn}{\nonumber \\}
\newcommand{\p}{\partial}
\newcommand{\pd}[1]{\frac {\partial} {\partial #1}}
\newcommand{\cd}{\nabla}
\newcommand{\cc}{$>$}
\newcommand{\bqa}{\begin{eqnarray}}
\newcommand{\eqa}{\end{eqnarray}}
\newcommand{\bqe}{\begin{equation}}
\newcommand{\eqe}{\end{equation}}
\newcommand{\bay}[1]{\left(\begin{array}{#1}}
\newcommand{\eay}{\end{array}\right)}
\newcommand{\eg}{\textit{e.g.} }
\newcommand{\ie}{\textit{i.e.}, }
\newcommand{\iv}[1]{{#1}^{-1}}
\newcommand{\at}[1]{{\Big|}_{#1}}
\newcommand{\zt}[1]{\texttt{#1}}
\newcommand{\non}{\nonumber}
\newcommand{\m}{\mu}
\def\xa{{m}}
\def\xA{{m}}
\def\xb{{\beta}}
\def\xB{{\Beta}}
\def\xd{{\delta}}
\def\xD{{\Delta}}
\def\xe{{\epsilon}}
\def\xE{{\Epsilon}}
\def\xve{{\varepsilon}}
\def\xg{{\gamma}}
\def\xG{{\Gamma}}
\def\xk{{\kappa}}
\def\xK{{\Kappa}}
\def\xl{{\lambda}}
\def\xL{{\Lambda}}
\def\xo{{\omega}}
\def\xO{{\Omega}}
\def\xvp{{\varphi}}
\def\xs{{\sigma}}
\def\xS{{\Sigma}}
\def\xt{{\theta}}
\def\xvt{{\vartheta}}
\def\xT{{\Theta}}
\def \Tr {{\rm Tr}}
\def\CA{{\cal A}}
\def\CC{{\cal C}}
\def\CD{{\cal D}}
\def\CE{{\cal E}}
\def\CF{{\cal F}}
\def\CH{{\cal H}}
\def\CJ{{\cal J}}
\def\CK{{\cal K}}
\def\CL{{\cal L}}
\def\CM{{\cal M}}
\def\CN{{\cal N}}
\def\CO{{\cal O}}
\def\CP{{\cal P}}
\def\CQ{{\cal Q}}
\def\CR{{\cal R}}
\def\CS{{\cal S}}
\def\CT{{\cal T}}
\def\CV{{\cal V}}
\def\CW{{\cal W}}
\def\CY{{\cal Y}}
\def\BC{\mathbb{C}}
\def\BR{\mathbb{R}}
\def\BZ{\mathbb{Z}}
\def\sA{\mathscr{A}}
\def\sB{\mathscr{B}}
\def\sF{\mathscr{F}}
\def\sG{\mathscr{G}}
\def\sH{\mathscr{H}}
\def\sJ{\mathscr{J}}
\def\sL{\mathscr{L}}
\def\sM{\mathscr{M}}
\def\sN{\mathscr{N}}
\def\sO{\mathscr{O}}
\def\sP{\mathscr{P}}
\def\sR{\mathscr{R}}
\def\sQ{\mathscr{Q}}
\def\sS{\mathscr{S}}
\def\sX{\mathscr{X}}

\def\slz{SL(2,Z)}
\def\slr{$SL(2,R)\times SL(2,R)$ }
\def\ads{${AdS}_5\times {S}^5$ }
\def\adst{${AdS}_3$ }
\def\sun{SU(N)}
\def\ad#1#2{{\frac \delta {\delta\sigma^{#1}} (#2)}}
\def\bqf{\bar Q_{\bar f}}
\def\nf{N_f}
\def\sunf{SU(N_f)}

\def\dcirc{{^\circ_\circ}}

\author{Morgan H. Lynch}
\email{morgan.lynch@snu.ac.kr}
\altaffiliation{New address: Max-Planck-Institut f\"{u}r Kernphysik, Saupfercheckweg 1, 69117 Heidelberg, Germany}
\affiliation{Center for Theoretical Physics,
Seoul National University, \\ Seoul 08826, Korea}
\author{Evgenii Ievlev}
\email{evgenii.ievlev@nu.edu.kz}
\altaffiliation[On leave of absence from: ]{National Research Center “Kurchatov Institute”, Petersburg Nuclear Physics
Institute, St.\;Petersburg 188300, Russia}
\affiliation{Physics Department \& Energetic Cosmos Laboratory, Nazarbayev University,\\
Astana 010000, Qazaqstan}
\affiliation{Almaty University of Power Engineering and Telecommunications,\\ 
Almaty 050013, Qazaqstan}
\author{Michael R. R. Good}
\email{michael.good@nu.edu.kz}
\affiliation{Physics Department \& Energetic Cosmos Laboratory, Nazarbayev University,\\
Astana 010000, Qazaqstan}
\affiliation{Leung Center for Cosmology $\&$ Particle Astrophysics,
National Taiwan University, \\ Taipei 10617, Taiwan}

\title{Accelerated electron thermometer: observation of 1D Planck radiation}

\date{\today}

\begin{abstract}
We report on the observation of thermal photons from an accelerated electron via examination of radiative beta decay of free neutrons measured by the RDK II collaboration. 
The emitted photon spectrum is shown to corroborate a thermal distribution consistent with the dynamical Casimir effect. 
Supported by a robust chi-squared statistic, we find the photons reside in a one-dimensional Planck spectrum with a temperature predicted by the moving mirror model.

\end{abstract}

\maketitle

\tableofcontents
\section{Introduction}
 
Thermodynamic particle production, see e.g. \cite{Parker:1968mv, hawking1974black, Unruh:1976db}, discovered in the framework of quantum field theory in curved spacetime \cite{Parker:2009uva, Birrell:1982ix, Fabbri} is an emerging experimental science in both analog and fundamental systems. In analog settings, e.g. \cite{Unruh:1980cg}, experimental signatures are present for the major particle production mechanisms, e.g. the Parker \cite{Steinhauer:2021fhb}, Hawking \cite{Weinfurtner:2010nu}, and Fulling-Davies-Unruh effects \cite{Hu:2018psq}, as well as the dynamical Casimir effect (DCE) \cite{faccio}. The surprising robustness of these phenomena even applies to seemingly disparate systems such as rear-end collisions in traffic jams \cite{de2022black}. Moreover, quantum metrology is now on the cusp of bringing some of the most precise measurement apparatus online, utilizing these very tenets in their operation \cite{ivette}. The widespread ubiquity of thermodynamic quantum fluctuations, and their applications, also spreads beyond the analog regimes where their properties are being fully unraveled systematically. Fundamental systems also provide an avenue to explore such phenomena and thus solidify their thermodynamic interpretation as an intrinsic property of spacetime \cite{Bekenstein:1973ur, Jacobson:1995ab}.

High energy channeling radiation experiments at CERN-NA63 \cite{Wistisen:2017pgr} have indeed already opened the door to explore the Unruh effect and establish it as a fundamental property of spacetime \cite{Lynch:2019hmk}. There, the extremely large accelerations produced by the radiation reaction, or recoil, have given us confirmation that for accelerated observers, the quantum vacuum appears as an effervescent thermal bath at the Fulling-Davies-Unruh (FDU) temperature \cite{Fulling:1972md,Davies:1974th, Unruh:1976db}. It also appears that stimulated Hawking radiation from astrophysical black holes may be on the cusp of current detector sensitivities \cite{Abedi:2021tti}. Gravitational wave echoes could provide a unique window into this phenomenon by stimulating the emission of Hawking radiation and probing the nature of event horizons \cite{Abedi:2016hgu}. Finally, precision cosmology may even offer an avenue to explore Parker's effect \cite{parker1}. Multiple avenues are already being explored, such as anisotropies in the cosmic microwave background \cite{Agullo:2010ws}.

The DCE \cite{moore1970quantum,Wilson:2019ago} is a closely related fundamental phenomenon, which yields thermalized particle production under acceleration and can be viewed as an analog of black hole evaporation.  Often celebrated as the ``moving mirror model" \cite{Davies:1976hi,Davies:1977yv}, it involves a rapidly accelerating perfect reflector that emits particles that can be Planck distributed at a temperature proportional to the acceleration scale \cite{Davies:1974th}. Like the Unruh effect, which is characterized by interactions with thermalized quantum fluctuations \cite{Cozzella:2017ckb}, the accelerating boundary of the DCE brings those thermalized quantum fluctuations on-shell and emits them as radiation. Observational efforts are being deployed with the goal of direct detection \cite{Chen:2015bcg,Chen:2020sir} of relativistic moving mirror radiation and creation \cite{Brown2015}, and they complement the small but growing accumulation of observations of the dynamical Casimir effect (see references in \cite{Dodonov:2020eto}). Importantly, international efforts like the Analog Black Hole Evaporation via Lasers (AnaBHEL) experiment \cite{AnaBHEL:2022sri} corroborate the accelerating relativistic moving mirror as a viable avenue to probe the spectral physics of quantum vacuum radiation. Surprisingly, the extreme accelerations experienced by an electron during the process of radiative neutron beta decay \cite{nico,RDKII:2016lpd} also provides a robust system to examine the 3+1 dimensional electron dual of the 1+1 dimensional scalar \textit{thermalized} DCE \cite{Ritus:2022bph}, i.e. the moving mirror model.  

This leads to the notion that analogs of black hole evaporation may also, in fact, be found by examining radioactive decay. In the sense that a black hole will emit radiation and lower its mass, an unstable atom also decays similarly; i.e. radiation is emitted, and the mass is lowered. Although the exact mechanism of decay may differ, e.g. Hawking radiation as opposed to electroweak decay, there is evidence that nuclear decay, in particular beta decay, can be viewed as an analog of black hole evaporation. Neutron decay under acceleration has also been proposed to probe the closely related Unruh effect \cite{Matsas:1999jx}. In the black hole scenario, and under the moving mirror formalism, the mirror boundary acts as the origin of spherical coordinates in a higher-dimensional context. The mirror's acceleration parameter plays the role of the black hole surface gravity. For instance, in each of the three scenarios, the Schwarzschild \cite{Good:2016oey}, Reissner-Nordstr\"om \cite{good2020particle}, and Kerr \cite{Good:2020fjz} spacetimes, there is an exact radiative and thermal correspondence between the mirror and black hole particle production. 

Curiously, radiative beta decay provides an ideal setting to investigate thermality and, as it turns out when treated as moving point charge radiation, has a correspondence to the moving mirror model (as an instance of the DCE), which has a long history as an analog system used to study the thermal radiation from an evaporating black hole. For a visual of the electron, mirror, and black hole, see Fig. \ref{fig:bh}. Using beta decay is a new approach to examining characteristics of thermalized fluctuations in the quantum vacuum. Moreover, within the framework of widely recognized and carefully monitored radiative processes in the field of particle physics, beta decay offers an additional experimental opportunity to explore how these fluctuations arise in general when quantum fields are influenced by external circumstances (i.e. QFTCS, e.g. \cite{Birrell:1982ix,Parker:2009uva,Fabbri}).

The discovery of a potential link between the electron of radiative beta decay and the moving mirror of the DCE was initially hinted at through the form of radiation reaction pointed out by both Unruh-Wald \cite{Unruh:1982ic} and Ford-Vilenkin \cite{Ford:1982ct} in 1982. This connection was further strengthened in 1995 when Nikishov-Ritus \cite{Nikishov:1995qs} established a formal link through particle count. Ritus \cite{Ritus:1999eu,Ritus:2002rq,Ritus:2003wu,Ritus:2022bph} later provided additional development on the Bogolyubov-current connection. On top of this, the correspondence between the two phenomena was confirmed with the derivation of an analog Larmor power by Zhakenuly et al \cite{Zhakenuly:2021pfm}. Explicit solutions have exploited the electon-mirror connection; e.g. see the demonstration between radiation power loss and kinetic power loss for an electron that approaches the speed of light \cite{good2022extreme} or the treatment of an electron as a tiny mirror for a trajectory that asymptotically approaches a constant velocity \cite{Good:2023hsv}.  The trajectory solution we focus on in this paper was first derived from thermal emission by an analog black hole, written as a Memorial for Kerson Huang \cite{Good:2016yht}. The connection of that mirror trajectory to the electron's motion was recently advanced in \cite{Good:2022eub}. 

Here we report on the observation of thermal radiation, in the form of a 1 dimensional Planck distribution, emitted via the radiative decay mode of the free neutron. Moreover, the temperature of this radiation is predicted by the electron-mirror dual description of neutron beta decay. Building upon the robust timeline of well-established developments in the theory of this duality, we now have an experiment to analyze that can potentially confirm or disconfirm the interpretation of the electron's emission as being thermal in accordance with a perfectly reflecting accelerated moving mirror. 

We examine the photon spectra produced by the radiative decay mode of the free neutron using the ``product log of an exponential" (PROEX) trajectory of \cite{Good:2016yht} and the temperature predicted by the Bogoliubov connection of the associated acceleration radiation \cite{Good:2022eub}; this electron-mirror symmetry has been developed throughout the past half-century but has, to date, remained untested. Indeed, we find the temperature of the radiation emitted in complete agreement with that predicted by the duality using the Bogoliubov mechanism and present an excellent chi-squared statistic matching the theory to the experimental data without fitting, i.e. the theory is exact. Moreover, radiation reaction in the high-frequency tail is confirmed via the same parameter-free analysis. Thus, we present evidence of thermal photons emitted by an accelerating electron, dual to a moving mirror, from the radiative beta decay mode of the free neutron. In the next 
subsection, Sec. \ref{sec:mec}, we comment on the role of $\hbar$ for temperature in SI units, but otherwise, we use natural units: $	\hbar = \mu_0 = c = k_B = 1$, $ e^2 \approx 0.0917012$.

\section{Theoretical background}\label{sec:background}

\subsection{Mirror-electron correspondence}\label{sec:mec}

\begin{figure}
\includegraphics[width=0.98\columnwidth]{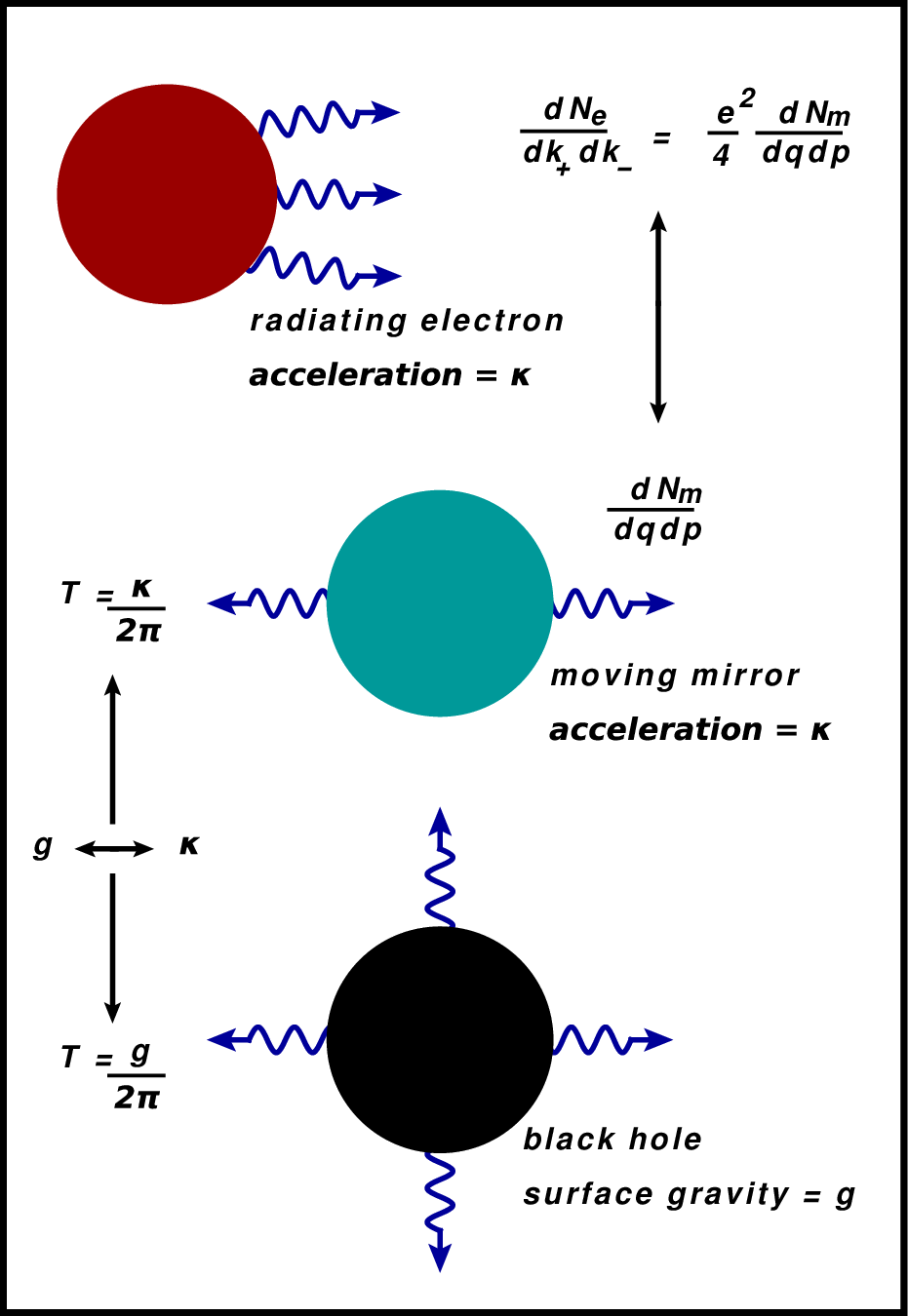}
\caption{
		\label{fig:bh}
		A heuristic diagram demonstrating (1) the forward directional beaming of classical radiation spectrum from an accelerating election, (2) the bi-directional quantum radiation spectrum of a moving mirror in 1+1 dimensions, and (3) the dual, $\kappa \leftrightarrow g$, temperature of the radial evaporation from a black hole. 
		For discussion, see Sec.~\ref{sec:mec}.
	 }
\end{figure}

In consideration of a moving mirror that reflects, or emits, a massless scalar field $\Phi(z)$ in 1+1 dimensions, we define the mirror by the Dirichlet boundary condition $\square \Phi (z) = 0$ on the mirror surface \cite{DeWitt:1975ys,Davies:1976hi,Davies:1977yv}. 
This scalar field can be expanded in modes, $\phi$, which satisfy the light cone wave equation, $\partial_{u}\partial_{v}\phi=0$. Here, we have defined the coordinates, $u = t - z$ and $v=t+z$. Our two sets of solutions to our wave equation then define our early time, i.e. in states, and late time, i.e. out states.
\bqa
\phi^{\textrm{in}}_{p} &=& \frac{1}{\sqrt{4 \pi p}} \lbk e^{- i p v} - e^{- i p v(u)} \rbk, \non \\
\phi^{\textrm{out}}_{q} &=& \frac{1}{\sqrt{4 \pi q}} \lbk e^{-i q u(v)} - e^{- i q u} \rbk.
\label{in_out_modes}
\eqa
The functions $v(u)$ and $u(v)$ characterize the mirror trajectory in light cone coordinates and enforce the boundary condition $\phi(u,v)=0$ at the mirror's location. The in and out-state momenta are labeled by $p$ and $q$, respectively. Since the in-modes form a complete set, one can expand the out-modes in terms of them:
\begin{equation}
	\phi^{\textrm{out}}_{q} = \int\limits_0^\infty \diff{p} \lbk  \alpha_{pq} \phi^{\textrm{in}}_{p} + \beta_{pq} \phi^{\textrm{in} \ast}_{p}\rbk .
\end{equation}
The quantities $\alpha_{pq}$ and $\beta_{pq}$ are nothing but the Bogoliubov coefficients.
Using the orthonormal property of the modes Eq.~\eqref{in_out_modes}, one can write down the formula for these coefficients in terms of respective scalar products. Inverting the Bogoliubov expansion then yields our coefficients,
\bqa
\alpha_{pq}&=& i\int dx \lbk \phi^{\textrm{in} \ast}_{p} \frac{\overleftrightarrow{\partial}}{\partial t} \phi^{\textrm{out} }_{q} \rbk , \non \\
\beta^{\ast}_{pq}&=& i\int dx     \lbk \phi^{\textrm{in} \ast}_{p} \frac{\overleftrightarrow{\partial}}{\partial t} \phi^{\textrm{out} \ast }_{q} \rbk .
\eqa
Utilizing the explicit form of our mode decomposition then yields the trajectory-dependent Bogoliubov coefficients. Hence,
\bqa
\alpha_{pq}&=& \frac{1}{2\pi}\sqrt{\frac{p}{q}}\int\limits_{-\infty}^\infty \diff{v} e^{ipv+iq u(v)}, \non \\
\beta_{pq}&=& \frac{1}{2\pi}\sqrt{\frac{q}{p}}\int\limits_{-\infty}^\infty \diff{u} e^{iqu+ip v(u)}.\label{beta_formula}
\eqa

The differential spectrum, $\diff{N^{\rightarrow}_\textrm{m}}$ of emitted particles to an observer on the right, produced by mixing the positive and negative frequency modes of our in and out states, is determined by the $\beta$ coefficient. As such, for an arbitrary trajectory, the particle creation is
\bqa
\diff{N^{\rightarrow}_{\textrm{m}}} &=& |\beta_{p q}|^{2}\diff{p}\diff{q}, \non \\
&=& \frac{\diff{p}\diff{q}}{(2 \pi)^{2}} \frac{q}{p} \left| \int_{-\infty}^{\infty} \diff{u} e^{iqu+ip v(u)} \right|^{2}. \label{dn_mir}
\eqa 

As we shall see, the above 1+1 dimensional quantum moving mirror spectrum is dual to that of 3+1 dimensional radiating electron. In this regard, let us now turn to radiation in classical electrodynamics. In consideration of a classical charge propagating along the trajectory, $x_{\textrm{tr}}(t)$, the subsequent radiation spectrum, $\diff{N_e}$, determined by the Fourier current density, $j_{\alpha}(k)$, is given by,
\bqa
\diff{N_{e}} &=& \left| j_{\alpha}(k)  \right|^2 \frac{\diff^3 {k}}{16 \pi^3 k_{0}}, \non \\
j_\alpha (k) &=& e\, \int\limits_{- \infty}^{\infty} \diff{t} \, \dv{x_{\alpha}(t)}{t} \, e^{-i k_{\mu}x_{\textrm{tr}}^{\mu}(t) }. \label{fourier_current}
\eqa
Here, $k_{0} = |k|$ and $t$ is the coordinate time along the trajectory. We define the light cone coordinate trajectory in retarded time, $u=x_{-}(t) = t -x_{\textrm{tr}}(t)$, and advanced time $v=x_{+}(t)= t + x_{\textrm{tr}}(t)$, to parameterize the system.  With a rectilinear trajectory along the $z-$axis, our current density will then be comprised of two components,

\bqa
j_{0} &=& -e \int_{-\infty}^{\infty} \diff{u} \frac{k_{z}}{k_{-}} e^{\frac{i}{2}(k_{+}u + k_{-}v(u))}, \non \\
j_{z} &=& -e \int_{-\infty}^{\infty} \diff{u} \frac{k_{0}}{k_{-}} e^{\frac{i}{2}(k_{+}u + k_{-}v(u))}.
\eqa
Here we have defined the light cone momenta $k_{\pm}=k_{0} \pm k_{z} = \omega(1\pm \cos\theta)$. Then, by making use of the Jacobian, $\frac{\diff^3{k}}{k_{0}} = \frac{1}{2}\diff{k_+}\diff{k_-}\diff{\phi} \rightarrow \pi \diff{k_+}\diff{k_-}$, with $0\leq k_{\pm} \leq \infty$, we have our electron spectrum. Hence,

\bqe
\diff{N_{e}} =  \frac{\diff{k_+}\diff{k_-}}{16 \pi^2} \frac{k_{+}}{k_{-}} \left| e \int_{-\infty}^{\infty} \diff{u} e^{\frac{i}{2}(k_{+}u + k_{-}v(u))}  \right|^2. \label{dn_ele}
\eqe 

We now see that both the mirror's scalars, Eq.~\eqref{dn_mir}, and electron's photons, Eq.~\eqref{dn_ele}, have a similar form.  These two expressions can be mapped via the light cone frequencies of the photons emitted by the electron to the light cone frequencies of the scalars emitted by the mirror. Thus when we have $p = \frac{k_{-}}{2}$ and $q=\frac{k_{+}}{2}$, along with scaling of the electron spectrum by $e^2$, our two spectra match identically:
\bqe
 \frac{\diff{N^\rightarrow_{\textrm{m}}}}{\diff{p}\diff{q}} = \frac{4}{e^2} \frac{\diff{N_{e}}}{\diff{k_+}\diff{k_-}}. \label{dn}
\eqe
The above expression also maps between the electron current and the Bogoliubov mirror coefficient \cite{Ievlev:2023inj}, 
\bqe
 |\beta_{pq}|^2 = \frac{|j_{\alpha}(k)|^2}{(2 \pi e)^2}.\label{betaJ}
\eqe
This duality manifests itself in more places than just the particle spectra. The total energy radiated by a mirror and an electron also obeys a similar relation. In terms of $\beta_{pq}$, the total energy radiated on both sides of the mirror is given by the formula
\begin{equation}
	E_{\textrm{mirror}} = \int\limits_0^\infty \diff{p} \int\limits_0^\infty \diff{q} \ (p + q)\; |\beta_{pq}|^2  \ . 
\label{Etot_betas_1}
\end{equation}

Likewise, the electromagnetic radiation from that electron can be calculated provided its trajectory $z(t)$ and velocity $\dot{z}(t)$ are known. 
In particular, for an electron moving in a rectilinear fashion along the $z$-axis, the spectral energy density of the radiation is given by (see Eq. (14.70) of Jackson \cite{Jackson:490457} or Eq. (23.89) of Zangwill \cite{Zangwill:1507229}, and \cite{Ievlev:2023inj} for a discussion)
\begin{equation}
	\frac{\diff I (\omega, \theta)}{\diff \Omega} = 
		\frac{\omega^2}{16 \pi^3} \sin^2\theta \, \abs{   j_z(\omega, k_z ) }^2 \,.
\label{I_Omega}
\end{equation}
Here, $\theta$ is the azimuthal angle, and $j_z(\omega, k_z )$ is the Fourier transform of the current density.
The total radiated energy can be calculated by integrating the spectrum,
\begin{equation}
	E_{\textrm{electron}}
		= \int\limits_0^\infty \diff{\omega} \int \diff{\Omega}\, \frac{\diff I(\omega,\theta)}{\diff \Omega} \,.
\label{Etot_pointcharge}
\end{equation} 


What is the connection between these two energies, Eq.~(\ref{Etot_pointcharge}) and Eq.~(\ref{Etot_betas_1})? It turns out they only differ by a factor of $e^2$.  Hints of this duality originated in \cite{Unruh:1982ic,Ford:1982ct,Nikishov:1995qs} and was further worked out in detail in \cite{Ievlev:2023inj,Ievlev:2023bzk}.
Let us briefly review a few more aspects of it here.

The mirror's radiation is fully described by the $\beta_{pq}$ coefficients with the two frequencies $p$ and $q$.
On the other hand, the electron's radiation is described by the spectral distribution $\diff I / \diff \Omega$ that depends on the frequency $\omega$ and the polar angle $\theta$. (Since the electron is moving in a rectilinear fashion, the system is axially symmetric, and the spectrum does not depend on the azimuthal angle $\phi$.)
The correspondence between these two systems can be formulated with the help of the mapping

\begin{equation}
    p + q = \omega \,, \quad
    p - q = \omega \cos\theta \ .
\label{pq_definition_2}
\end{equation}
Using this map as an integral transformation on the $\beta$ Bogolubov coefficients Eq.~\eqref{beta_formula} (see \cite{Ievlev:2023bzk} for a detailed derivation), one can show that they correspond to the Fourier-transformed electron current,
\begin{equation}
	|\beta_{p q}| = \frac{\sin\theta }{2 \pi e  }\ j_z (\omega, \omega\cos\theta)\ . 
\label{recipe-betaR}
\end{equation}
The frequencies $p,q$ on the l.h.s. are related to $\omega$ and $\cos\theta$ on the r.h.s. via Eq.~\eqref{pq_definition_2}.

The same transformation can be performed on the energy integral Eq.~\eqref{Etot_betas_1}. One finds, as promised, that
\begin{equation}
	E_{\textrm{mirror}} = \frac{1}{e^2} E_{\textrm{electron}} \,.
\label{recipe-energy}
\end{equation}
We stress that the formulas Eq.~\eqref{recipe-betaR} and Eq.~\eqref{recipe-energy} are quite general and work for any asymptotically inertial trajectory.
Several more noteworthy quantities are also related: radiation spectra,
\begin{equation}
	\frac{1}{4\pi} |\beta_{pq}|^2 = \frac{1}{e^2 \omega^2} \, \frac{\diff{I}}{\diff{\Omega}}(\omega,\cos\theta) \,,
\label{I_beta_recipe}
\end{equation}
the Larmor-Li\'enard formula for power radiated \cite{Zhakenuly:2021pfm},
\be 
P_{\textrm{mirror}} = \frac{1}{e^2}P_{\textrm{electron}} = \frac{ \alpha^2}{6\pi}\,,
\ee
and Lorentz-Dirac self-force magnitudes \cite{Myrzakul:2021bgj},
\begin{equation}
F_{\textrm{mirror}} = \frac{1}{e^2} F_{\textrm{electron}} = \frac{ \alpha'}{6\pi}\,.
\end{equation}
Here $\alpha$ is the proper acceleration of the mirror or electron, and the prime is a derivative with respect to the proper time. A caricature of the mirror's bi-directional radiation and the electron's forward-beamed radiation is in Fig.~\ref{fig:bh}.

\begin{table}[ht] 
\centering
\begin{tabular}{|>{\centering\arraybackslash}m{2cm}|>{\centering\arraybackslash}m{2cm}|>{\centering\arraybackslash}m{2cm}|>{\centering\arraybackslash}m{2cm}|}
  & Black Hole & Mirror &  Electron   \\
\hline\hline
Quantum $\hbar$ & $\checkmark$ & $\checkmark$  & \xmark \\ 
Gravity $G$ & $\checkmark$ & \xmark & \xmark 
\end{tabular} \\ 
\caption{A summary of the physical contexts.  The radiation from the electron is classical.  In the mirror-electron pictures, the acceleration parameter $\kappa$ corresponds to the surface gravity in the black hole picture $\kappa\leftrightarrow c^4/(4GM)$. In the electron picture, $\hbar$ is replaced by the electron charge: $\hbar \leftrightarrow \mu_0 c e^2$. See the exception for the appearance of $\hbar$ in Kelvin acceleration-temperature, Eq.~(\ref{sihbarkelvin}) and Eq.~(\ref{tempacckelvin}).} 
\label{role_of_parameters} 
\end{table}

The connection outlined here is a duality between a quantum field theory in 1+1d (moving mirror) and a classical theory in 3+1d (electron), cf. Table~\ref{role_of_parameters} and \ref{dualitytable}.
If we were to switch from natural to SI units, we would see that the proportionality coefficient between the total energies in Eq.~\eqref{recipe-energy} is in fact
\begin{equation}
	\frac{ \hbar }{ \mu_0 c e^2 } \approx 10.91 \,,
\end{equation}
where $c$ is the speed of light and $\mu_0$ is the vacuum magnetic permeability, see \cite{Ievlev:2022emz} for an extended discussion.

Let us consider the temperature of radiation in this context.  It would seem that the temperature of the mirror's quantum radiation should contain $\hbar$, while the electron's temperature should be expressible in purely classical terms. However, contrary to intuition, classical radiation temperature (measured in Kelvin) requires Planck’s constant.  This has been the case since 2009 because, by fiat definition \cite{SI}, Planck’s constant defines the SI unit for temperature. 

This is interesting because the temperature we report from the one-dimensional Planck distribution is classical.  For orientation, consider that the Kelvin temperature scale used to be defined in terms of the triple point of water. However, now \cite{SI} that the Kelvin scale is defined in terms of Boltzmann's constant, caesium-133 hyperfine transition frequency, and Planck's constant, 
\be 1\; \textrm{K} = \frac{2\pi \times 1.380649 \times 10^{-23}}{
(6.62607015 \times 10^{-34}) (9192631770)}\frac{\Delta \nu_{\textrm{Cs}} \hbar}{k_B}, \label{sihbarkelvin}\ee
it is important to realize the appearance of $\hbar$ in the  acceleration-temperature SI equation, e.g.
\be T = \frac{\hbar a}{2\pi c k_B},\label{tempacckelvin}\ee
where $a$ is some acceleration, does not signify the effect is quantum. In the case of classical thermal radiation from an accelerating electron, $\hbar$ only means that the SI unit Kelvin scale is being used. See Table \ref{dualitytable} for a chart.
\begingroup
    \renewcommand{\arraystretch}{2}
\begin{table}[ht] 
\centering
\begin{tabular}{|>{\centering\arraybackslash}m{3.75cm}|>{\centering\arraybackslash}m{3.75cm}|}

\hline
 & Electron-Mirror \\
\hline\hline
Power  & $P_{\textrm{electron}} = e^2 P_{\textrm{mirror}}$ \\\hline
Force  & $F_{\textrm{electron}} = e^2 F_{\textrm{mirror}}$ \\\hline
Energy & $E_{\textrm{electron}} = e^2 E_{\textrm{mirror}}$ \\\hline

Particles & $N_{\textrm{electron}} = e^2 N^{\rightarrow}_{\textrm{mirror}}$ \\\hline
Temperature (Stoney) & $T_{\textrm{electron}} = e^2 T_{\textrm{mirror}}$ \\
Temperature (Kelvin) & $T_{\textrm{electron}} = T_{\textrm{mirror}}$ \\
\hline
\end{tabular}
\caption{A chart of some of the quantities in the electron-mirror duality that are proportional.  We have used natural units, $\hbar = \mu_0 = c = k_B = 1$, $ e^2 \approx 0.0917$. The row for particles count one side of the mirror: $2 N^{\rightarrow}_m = N^{\leftrightarrow}_m$. The row for temperature demonstrates the sensitivity to scale as described for Eq.~(\ref{tempacckelvin}). The power, force, and energy radiated by the mirror are about 11 times more than those radiated by the electron for a given accelerated trajectory.} 
\label{dualitytable} 
\end{table} 
\endgroup

To close this section, we note that the 1+1d moving mirror model has much in common with evaporating black holes in 3+1 dimensional spacetime \cite{DeWitt:1975ys,Davies:1976hi,Davies:1977yv}.
Particle production by a collapsing star which forms a black hole (Hawking radiation), is caused by a changing spacetime geometry.
Particle production from the mirror is also caused by a changing geometry, although in a much simpler mechanism: expansion or contraction of flat space by the moving boundary condition.
In \cite{1993bhmw.conf....1W}, this has been brought into a quantitative form, and a mirror's trajectory was found that corresponds to a null dust shell collapsing to form a Schwarzschild black hole.

On both sides, characteristic parameters have the dimension of acceleration: the surface gravity $g$ for the black hole and the acceleration scale $\kappa$ for the mirror's trajectory. The mathematically stated correspondence means that the beta Bogoliubov coefficients, and thus their spectra, for the mirror and the black hole are the same upon identification of $g$ and $\kappa$. This implies that the thermal radiation temperatures for the mirror and black hole radiation are also the same:
\begin{equation}
	T_\text{mir} = \frac{\kappa}{2 \pi} \quad
	\Leftrightarrow \quad
	T_\text{BH} = \frac{g}{ 2 \pi } \,.
\end{equation}
Through this black hole-mirror correspondence, we obtain the thermal dynamical Casimir effect. Then, through the electron-mirror correspondence, we are able to explore this thermality via the strongly accelerated systems of nuclear decay. This longstanding duality between moving mirror temperature and black hole temperature has yielded tremendous insight into the nature of the two quantum systems under acceleration and/or gravity, respectively. For systems such as the Unruh effect \cite{Lynch:2019hmk, Unruh:1976db} or issues like the information problem \cite{Almheiri:2019psf, Engelhardt:2014gca,2023arXiv230314642L}, significant insights have been gained by understanding their underlying connections. The moving mirror model and its electromagnetic counterpart have an extremely close connection to each other, and, as we shall see, the experimental setting of beta decay is an area that can be used to explore this relationship. Taken together, Table~\ref{role_of_parameters} and \ref{dualitytable}, and Fig.~\ref{fig:bh} provide some overview of the interplay between the mirror,  electron, and black hole correspondence.

\subsection{Thermal particle emission}

Now, having discussed radiation spectra for the mirror and the electron, let us turn to the particular problem of electrons in beta decay.
Here we are going to propose a specific model describing such electrons and their photon emission.
The mirror-electron duality will help us to determine the temperature of the photon radiation.

\subsubsection{Temperature from the moving mirror}
\label{mirror_motivation}

Let us motivate a particular worldline trajectory for our model. In determining the trajectory of interest, we are motivated to reproduce the total photon energy emitted \cite{Good:2016yht} during beta decay. This is the most important connection linking the theoretical analysis to the experiment. Given a typical velocity Heaviside theta function trajectory, the resulting spectrum is independent of frequency e.g. \cite{Zangwill:1507229}. In contrast, the continuous PROEX worldline \cite{Good:2016yht}, results in a frequency-dependent spectrum \cite{Good:2022eub}, which results in the total energy emitted by the inner bremsstrahlung of radiative beta decay. 

Beside a frequency-dependent spectrum, the motivation for this choice of trajectory is additionally supported by the fact that given a final velocity $|\beta_{
\textrm{f}}| = s$, the trajectory has a  proper acceleration, $\mathcal{A}(t) = \kappa \beta \gamma^{3}(1-\beta/s)^2$ which is unique within the M\"obius group \cite{mobmir}. Here $\kappa$ is the acceleration scale. 

Consider an accelerated mirror trajectory following along the worldline of PROEX, whose Bogoliubov transformation between the early and late times of the mirror modes yields the overall spectrum \cite{Good:2022eub},

\bqe |\beta_{pq}|^2 = \frac{2 s^2 pq}{\pi  \kappa  (p+q) }\frac{ a^{-2} + b^{-2}}{ e^{2\pi (p+q)/\kappa}-1}. \label{fullspectrum}\eqe
Here $a= p(1+s) + q(1-s)$, $b=p(1-s)+q(1+s)$, and $q$ and $p$ are the frequencies of the in and out modes respectively. The first thing to note here is a thermal flux of scalar radiation from the mirror, which is indicated by the explicit Bose-Einstein distribution in the above expression. 

Let us make the thermality more explicit by considering high final speeds, $s\approx 1$, then using $q\gg p$, the high-frequency approximation \cite{Hawking:1974sw}.  To leading order, one gets 
\bqe 
|\beta_{pq}|^2 = \frac{1}{2 \pi  \kappa p} \frac{1}{e^{2\pi q/\kappa} -1}. \eqe
High final speeds and the low-frequency approximation, $q \ll p$ switches $p\leftrightarrow q$, which gives the widely agreed on result for thermal radiation \cite{Fulling_optics},
\bqe 
|\beta_{pq}|^2 = \frac{1}{2 \pi  \kappa q} \frac{1}{e^{2\pi p/\kappa} -1}. \label{beta} \eqe 
The temperature $T$ is set by the asymptotic scale $\kappa$ of the peel acceleration $\bar{\kappa}$ \cite{carlitz1987reflections,Ievlev:2023inj, Bianchi:2014qua,Barcelo:2010pj}. The relationship between the peel acceleration, $\bar{\kappa}$, and proper acceleration, $\mathcal{A}$, is given by $\bar{\kappa} = 2 \mathcal{A} e^{\eta}$ \cite{good2022extreme}, here $\eta = \tanh^{-1} \beta$ is the rapidity along the worldline. In the ultra-relativistic limit and late times, the time-dependent local acceleration $\bar{\kappa}$ becomes the constant peel acceleration scale $\kappa$, i.e. $\bar{\kappa} = \kappa$.  

The spectral quantity Eq.~(\ref{beta}), found from the quantum scalar particle radiation emitted by the moving mirror implies a temperature \cite{Good:2022eub}:
\be T_{\textrm{FDU}} = \frac{\kappa}{2\pi}.\label{FDUkappatemp}\ee
By matching the total energy emitted by the mirror to that of radiative beta decay, one fixes the acceleration scale, $\kappa$, to the range of frequencies present in the electromagnetic radiation   \cite{Good:2022eub},
\bqe
\kappa = \frac{12}{\pi} \Delta_\omega \,, 
\label{kappa}
\eqe 
Here, $\Delta_\omega$, is the bandwidth of the electron's radiation. This matching has ostensibly applied the electron-mirror duality \cite{Ritus:2022bph}, but it is not needed if one assumes the electron created during beta decay moves along PROEX worldline from the start \cite{Ievlev:2023inj}. Regardless, the temperature Eq.~(\ref{FDUkappatemp}) characterizes the photon spectrum from radiative beta decay.

The electron-mirror duality allows one to map from quantum to classical radiation and vice versa. This motivated the consideration of radiative beta decay from a classical point of view. In our case, the PROEX trajectory employed is also unique under a M\"obius transform, which further motivated an investigation of its classical and quantum radiation. The M\"obius transform symmetry is preserved on both sides of the duality. Exploiting the electron-mirror duality allows us to map from the quantum expressions of the moving mirror to the classical expressions for the moving point charge. For details and additional duality transformation formulas, see recent developments on the dual-transformation recipe in \cite{Ievlev:2023inj,Ievlev:2023bzk}.

In summary, the electron-mirror duality was initially employed to understand that a trajectory involving a thermal scalar-emitting mirror can be equivalently understood as an electron trajectory that emits thermal photons. However, in hindsight, there is no strict necessity to invoke this duality. Instead, one can directly propose the electron trajectory and proceed from that point. The total photon energy emitted by the electron is \cite{Ievlev:2023inj}
\be E = \frac{e^2}{24\pi }\left(\frac{\tanh^{-1}s}{s} - 1\right)\kappa.\label{electronenergy}\ee
This $\kappa$ acceleration scale is connected to the bandwidth because we know the total photon radiation emitted during beta decay, e.g. \cite{PhysRev.76.365},
\be E =\frac{e^2}{24\pi} \left(\frac{\tanh^{-1}s}{s} - 1\right)\frac{12}{\pi}\Delta_{\omega}, \label{bd_energy}\ee  
thus establishing Eq.~(\ref{kappa}). 

The non-uniform acceleration of the electron, $\mathcal{A}(t)$, and the classical computation of the Planck spectrum, $I(\omega)$, gives the temperature $T_{\textrm{FDU}}$ in Eq.~(\ref{FDUkappatemp}) which characterizes the classical acceleration temperature or `thermal Larmor radiation.' It is understood using a broad but precise definition of the connection between temperature and acceleration via a standard classical computation of thermal emission from an accelerating point charge.  Directly linking temperature to $\kappa$ without appealing to the quantum radiation from the moving mirror is found by computing the classical radiation from the electron via the spectrum $I(\omega)$, see Sec. \ref{spectrumsec}, Eq.~(\ref{Iw}).

\subsubsection{Electron radiation and photon count} \label{spectrumsec}

Now let us discuss the spectral characteristics of the radiation coming from an electron in the beta decay.
The PROEX trajectory discussed in Sec.~\ref{mirror_motivation} has been studied in the context of electrodynamics in \cite{Ievlev:2023inj}, see also the discussion in \cite{Ievlev:2022emz}. In a nutshell, one should first evaluate the spectral distribution Eq.~\eqref{I_Omega} on the PROEX trajectory, $ z(t) = s W\lb e^{\kappa t} \rb/\kappa$.
The calculation is carried out in \cite{Ievlev:2023inj} and is somewhat involved, so we present here only the result:
\begin{equation}
	\frac{\diff I(\omega)}{\diff \Omega}  = \frac{ e^2 s^2 \sin ^2\theta }{16 \pi^3 (1-s \cos \theta)^2}\frac{2\pi  \omega/\kappa }{e^{2 \pi   \omega/\kappa}-1} \,.
\label{dI_dOmega}
\end{equation}
One can check that the electron spectral distribution Eq.~\eqref{dI_dOmega} is indeed related to the beta Bogoliubov coefficients in Eq.~\eqref{fullspectrum} by the duality mapping Eq.~\eqref{pq_definition_2} and Eq.~\eqref{I_beta_recipe}.

Integrating over the solid angle yields the spectrum
\begin{equation}
	I(\omega, \kappa, s) = \frac{e^2}{ \pi \kappa } \left(\frac{\eta}{s}-1\right) \frac{ \omega }{e^{2\pi  \omega/\kappa}-1} \,.
\label{Iw}
\end{equation}
Here, $\eta = \tanh^{-1} s$ is the final rapidity, while $\kappa$ is an acceleration parameter. With Eq.~(\ref{kappa}), $\kappa = 12 \Delta_\omega / \pi $, we fix the photon bandwidth via the electron kinetic energy. 
Since an electron with an energy $E_\text{kin}$ can radiate a photon in the frequency range $\omega \in [0, E_\text{kin}]$, one has:
\begin{equation}
	\kappa = \frac{12}{\pi} E_\text{kin} \,, \quad T_{\textrm{FDU}} = \frac{6}{\pi^2}E_\text{kin}.
\label{kappa_from_E}
\end{equation}

Note that while the explicit form of the Planck factor in Eq.~\eqref{Iw} corroborates the temperature expression in terms of $\kappa$, it does not tell us what $\kappa$ is. The latter is determined by matching the total radiated energy in beta decay, Eq.~(\ref{bd_energy}), to the energy emitted by the accelerated electron, Eq.~(\ref{electronenergy}). 

The nature of this temperature can also be examined via the relativistic work-kinetic energy theorem. There we have $ma\Delta x = E_{\textrm{kin}}$. For an acceleration length on the order of the Compton wavelength of the electron, we see explicitly that $\kappa \sim E_{\textrm{kin}}$ (recall that $\kappa$ is the acceleration scale). Via the moving mirror analysis, the coefficient is fixed by the conservation of energy of the system. 

Now, a particle count formula is needed. Since one photon has the energy $\omega$ (recall that $\hbar=1$), and Eq.~\eqref{Iw} gives energy density, we can write down the particles per keV:
\begin{equation}
\begin{aligned}
	n(\omega, \kappa, s) 
		&= \frac{1}{\omega} I(\omega, \kappa, s) \\
		&= \frac{e^2}{ \pi \kappa } \left(\frac{\eta}{s}-1\right) \frac{ 1 }{e^{2\pi  \omega/\kappa}-1} 	\,,
\end{aligned}
\label{particle_count_general}
\end{equation}
where
\begin{equation}
	s = s(E_\text{kin}) \,, \quad
		\kappa = \kappa (E_\text{kin}) \,. \label{sANDk}
\end{equation}
Here, $\kappa (E_\text{kin})$ is determined by Eq.~\eqref{kappa_from_E}, while the final speed $s$ is given by Eq.~\eqref{s_from_E}.

The entirety of the photon spectrum is now fixed by the kinematics of the electron. We have indeed found a spectrum which is thermalized at a temperature which is determined by the conservation of energy of beta decay process in conjunction with the moving mirror analysis. However, since the kinetic energy is determined via the probability distribution, Eq.~(\ref{decay_Ekin_distr}), we must now turn to how we select the energy which goes into our analysis.

\subsubsection{Incorporating recoil}

First, due to the fact that we have the emission of photons whose energy is approaching the rest mass of the electron, we should also expect to see signatures of radiation reaction or recoil. Recent experiments by CERN-NA63 along with plasma wake field accelerators have provided the first observations of this recoil effect \cite{Wistisen:2017pgr,2018PhRvX...8a1020C, 2018PhRvX...8c1004P, 2019PhRvR...1c3014W, 2008LMaPh..83..305P}. Moreover, in the NA63 data set, thermalization due to the Unruh effect, is also present in the data \cite{Lynch:2019hmk, 2023arXiv230314642L}.

The inclusion of recoil in these thermalized systems can be accomplished via a chemical potential \cite{Lynch:2019hmk,2023arXiv230106772G}, $\m$, in the thermal distribution which encodes the recoil kinetic energy, $\m = \frac{\omega^2}{2m}$. With this chemical potential, we can include recoil throughout our entire analysis by utilizing the following modified version of the photon count Eq.~\eqref{particle_count_general},
\begin{equation}
	n_{\textrm{rr}}(\omega, \kappa, s) = \frac{e^2}{ 2 \pi^2 } \left(\frac{\eta}{s}-1\right) \frac{ 1  }{e^{2\pi  \lb \omega +\frac{\omega^2}{2m} \rb/\kappa}-1} \,.
\label{rr}
\end{equation}
This recoil-corrected photon count
comes directly from the quantum recoil correction to the Larmor formula. It has the overall effect of steepening the spectrum and, in principle, leads to a spectral cutoff similar to that of the photon cutoff, $\bar{n}(\omega)_\text{cut}$, in the electron count.

\subsection{Electrons in beta decay}
\label{sec:E_pick}

\begin{figure}
\includegraphics[width=0.98\columnwidth]{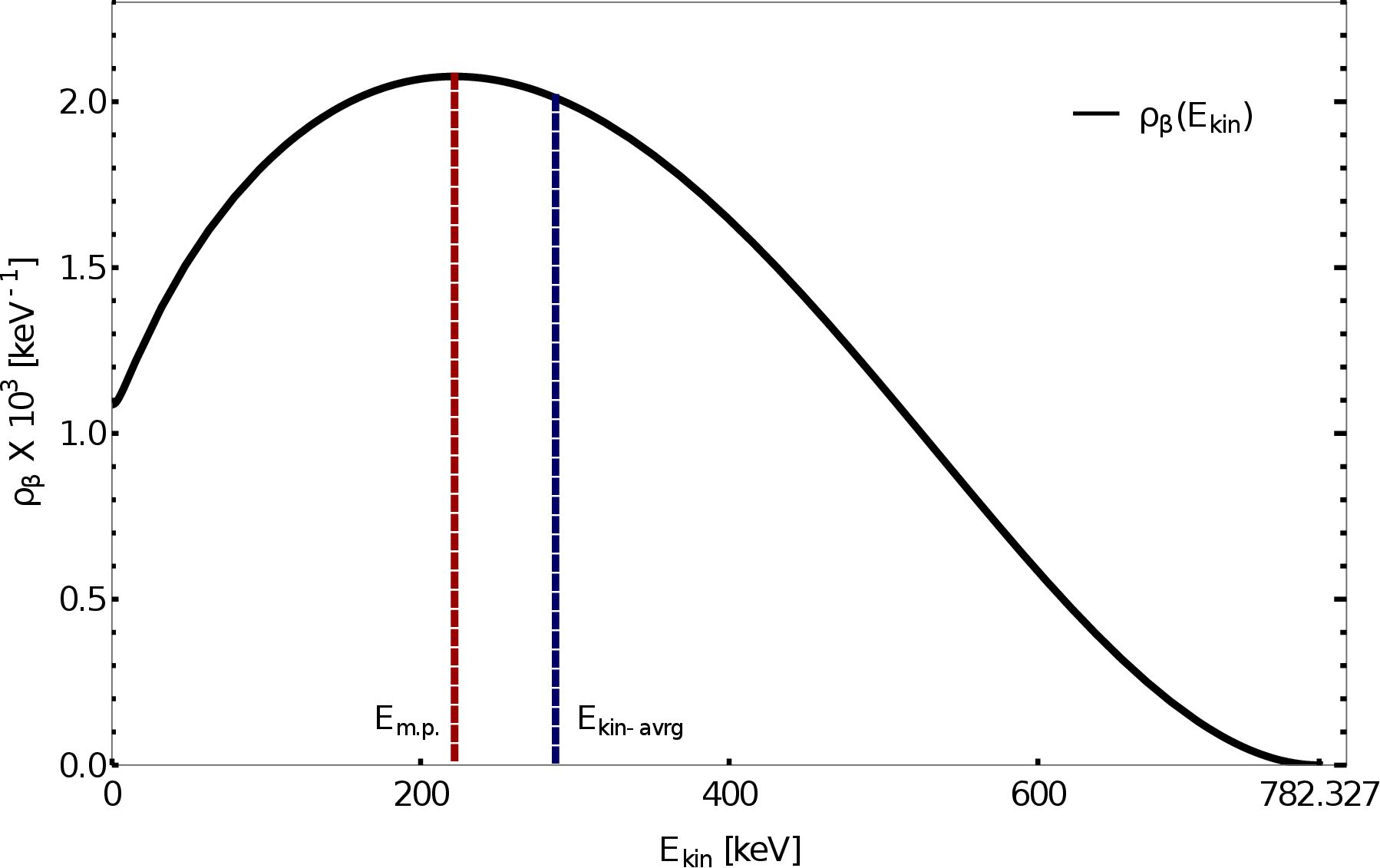}
\caption{
		\label{fig:fermi_distr}
		The probability distribution of the electron kinetic energy Eq.~\eqref{decay_Ekin_distr}. The most probable kinetic energy, $E_{m.p} = 221.98$ keV, and the average kinetic energy, $E_{kin-avrg} = 287.48$ keV, are utilized in our analysis, see section \ref{sec:E_pick} below. 
	 }
\end{figure}

For the following analysis, we will only be concerned with the free neutron $\beta^-$ decay \cite{RDKII:2016lpd}, i.e. when an electron is produced (one can also consider $\beta^+$ with positron production, but it is outside of the scope of the current paper).  The maximum possible kinetic energy of an electron in this decay is (we treat the neutrino as massless),
\begin{equation}
	E_\text{max} = m_n - m_p - m_e \approx 782.325 \text{ keV} \,.
\label{Emax}
\end{equation}
This corresponds to the speed,
\begin{equation}
	s_\text{max} \approx 0.918636 \,.
\end{equation}
However, the electrons that are actually produced in beta decay may have any kinetic energy $E_\text{kin}$ in the range 
$0 \leqslant E_\textrm{kin} \leqslant E_\textrm{max}$. 
The corresponding distribution can be derived from the Fermi golden rule. The resulting formula is given by (see e.g. \cite{Venkataramaiah_1985}),
\begin{equation}
	\rho_\beta (E_\text{kin}) = C \, F(E_\text{kin},Z) \, p \, (E_\text{kin} + m_e) \, (E_\text{max} - E_\text{kin})^2 \,.
\label{decay_Ekin_distr}
\end{equation}
Here, 
$Z$ is the charge number of the daughter nucleus (in our case $Z=1$), while  $C$ is a normalization constant determined by the condition
\begin{equation}
	\int\limits_0^{E_\text{max}} \diff E \ \rho_\beta (E) = 1 \,.
\end{equation}
The Fermi function, $F(E_\text{kin},Z)$, can be approximated by 
\begin{equation}
	F(E_\text{kin},Z) \approx \frac{2 \pi \zeta}{ 1 - e^{- 2 \pi \zeta} } \,, \quad
	\zeta = \frac{Z e^2}{ s} \,.
\label{F_approx}
\end{equation}
Here, $s = s(E_\text{kin})$ is the electron's final speed as a function of the kinetic energy,  
\begin{equation}
	s(E_\text{kin}) = \frac{ p(E_\text{kin}) }{ E_\text{kin} + m_e } \,.
\label{s_from_E}
\end{equation}
The relativistic momentum in Eq.~\eqref{decay_Ekin_distr} is given by
\begin{equation}
	p(E_\text{kin}) = \sqrt{ (E_\text{kin} + m_e)^2 - m_e^2 } \,.
\end{equation}
For reference, a plot of the kinetic energy probability distribution, Eq.~\eqref{decay_Ekin_distr}, is presented in Fig~\ref{fig:fermi_distr}.

In \cite{Venkataramaiah_1985} it was stated that the approximation Eq.~\eqref{F_approx} is valid for non-relativistic electrons. 
However, we checked that this approximation works well in our case too, with a relative error of $\sim 3 \cdot 10^{-4}$ in the energy range $E_\text{kin} \in [10^{-4} \text{ keV},E_\text{max}]$, which is more than enough for our purposes.

To compare with the experiment we need to pick some electron energy $E_\text{kin}$ and compute the corresponding quantities 
$s(E_\text{kin})$ and $\kappa (E_\text{kin})$, Eqs.~(\ref{sANDk}). We can motivate a few choices.

One choice is to pick the most probable energy $E_\text{m.p.}$ and use it for the particle count
\begin{equation}
	n_\text{m.p.}(\omega) = n(\omega, \kappa(E_\text{m.p.}), s(E_\text{m.p.})) \,.
\label{particle_count_mostprob}
\end{equation}
We find that the maximum of the distribution Eq.~\eqref{decay_Ekin_distr} in our case is located at
$E_\text{m.p.} = 221.98 \text{ keV}$.
Correspondingly, $s_\text{m.p.} = 0.716922$ and $\kappa_\text{m.p.} = 847.902$ keV.

Another choice one can take is to perform some kind of averaging with respect to the distribution Eq.~\eqref{decay_Ekin_distr}.
From the distribution Eq.~\eqref{decay_Ekin_distr} we can calculate the average energy:
$E_\text{avrg} = 287.482 \text{ keV}$.
Correspondingly, $s_\text{avrg} = 0.768406$ and $\kappa_\text{avrg} = 1098.1$ keV.
The particle count is then:
\begin{equation}
	n_\text{kin-avrg}(\omega) = n(\omega, \kappa(E_\text{avrg}), s(E_\text{avrg})) \,.
\label{particle_count_avrg_kin}
\end{equation}
Alternatively, one can compute the expectation value of the particle distribution itself. Namely, one introduces an averaged particle count
\begin{equation}
	\bar{n}(\omega) = \int\limits_{0}^{E_\text{max}} \, \diff E \ \rho_\beta (E) \, n(\omega, \kappa(E), s(E)) \,.
\label{particle_count_avrg_distr}
\end{equation}
A further reasonable modification can be made by crudely enforcing the energy conservation law: we require that an electron with an energy $E$ cannot possibly create a photon whose frequency exceeds $E$. This can be done by inserting a Heaviside step function:
\begin{equation}
	\bar{n}(\omega)_\text{cut} = \int\limits_{0}^{E_\text{max}} \, \diff E \ \theta( E - \omega) \, \rho_\beta (E) \, n(\omega, \kappa(E), s(E)) \,.
\label{particle_count_avrg_cut_distr}
\end{equation}
Below, we will compare these approaches.

\section{Comparing with experimental data}

\begin{figure*}[t]
	\subfloat[APD spectrum]{\label{fig:distr_APD}\includegraphics[width=0.98\columnwidth]{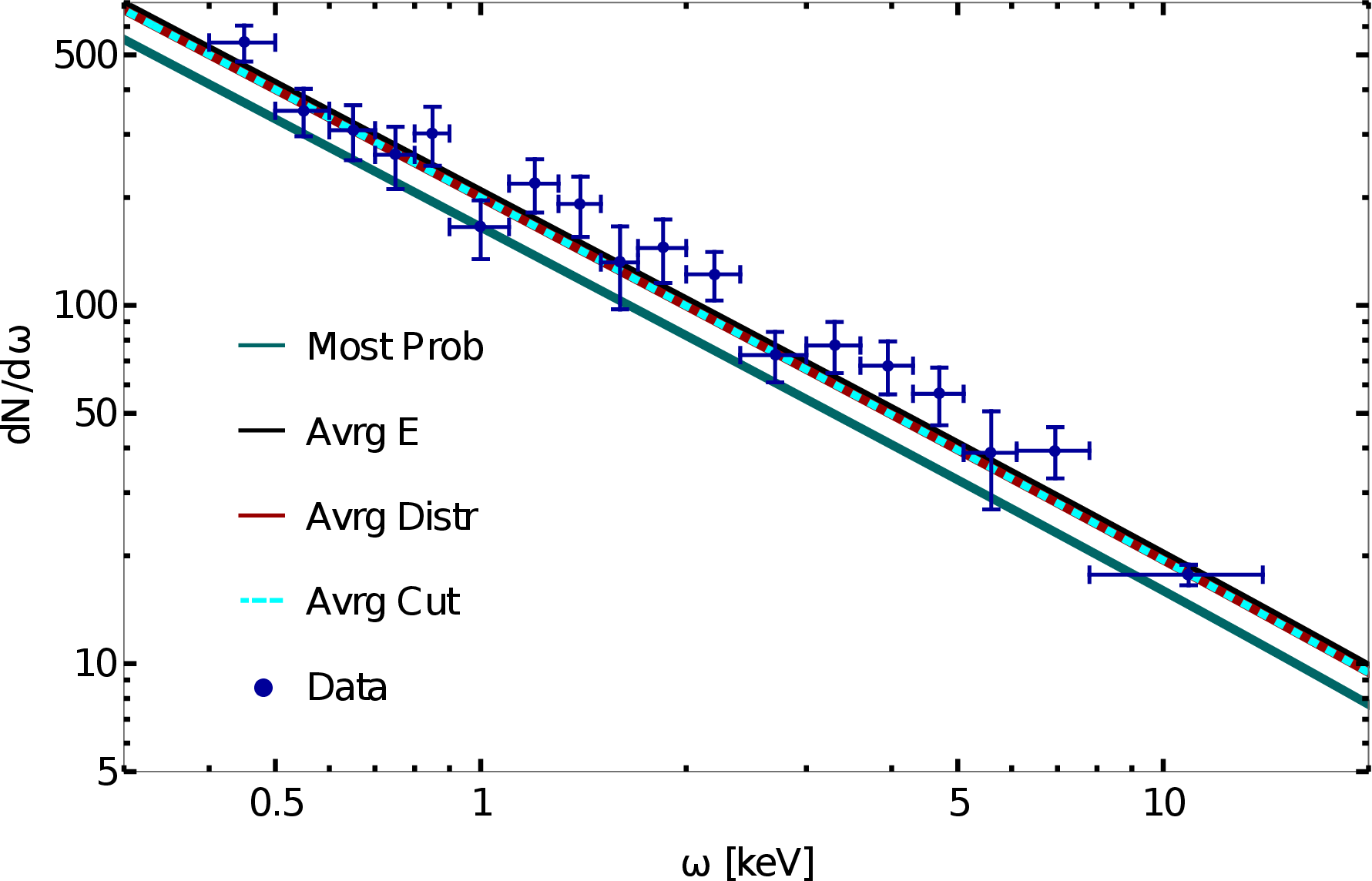}}
	\qquad
	\subfloat[BGO spectrum]{\label{fig:distr_BGO}\includegraphics[width=0.98\columnwidth]{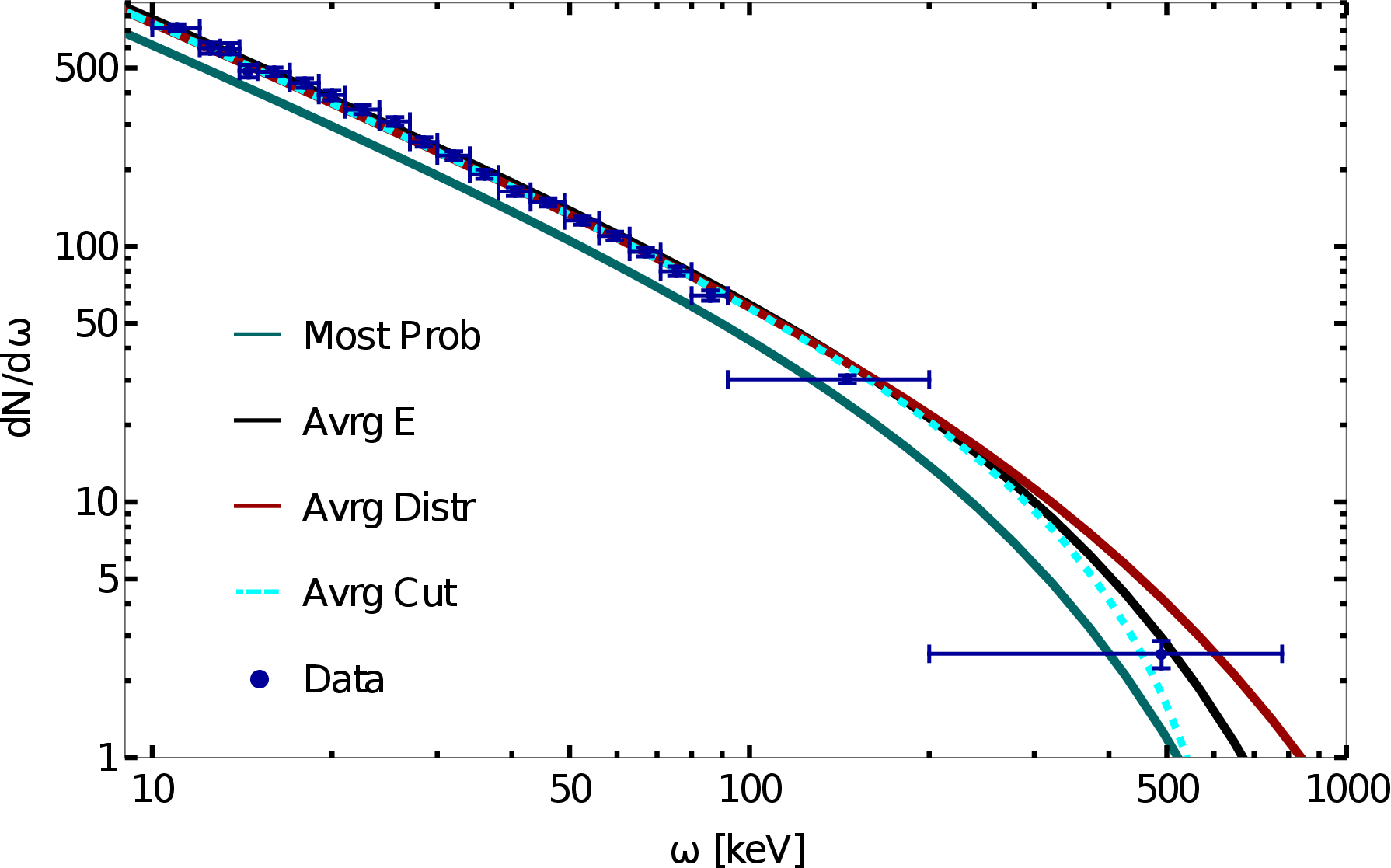}}
\caption{
		Particle distribution for the APD and BGO detectors.
		The blue dots are the RDK II experimental data.
		The curves are the results of our thermalized 1-D Planckian analysis (without fitting) of Eq.~\eqref{distr_to_compare}, without the recoil correction.
		Each of curve corresponds to a particular choice for the electron kinetic energy and thus particle distribution $n(\omega)$: most probable kinetic energy $E_{kin}$ Eq.~\eqref{particle_count_mostprob}, the average kinetic energy $E_{kin-avrg}$ Eq.~\eqref{particle_count_avrg_kin}, the average kinetic energy over the distribution Eq.~\eqref{particle_count_avrg_distr},
		the average kinetic energy over the distribution with the kinetic cutoff Eq.~\eqref{particle_count_avrg_cut_distr}, respectively. 
		\label{fig:distr}
	}
\end{figure*}

\begin{figure*}[t]
	\subfloat[APD bins]{\label{fig:distr_APD_bins}\includegraphics[width=0.98\columnwidth]{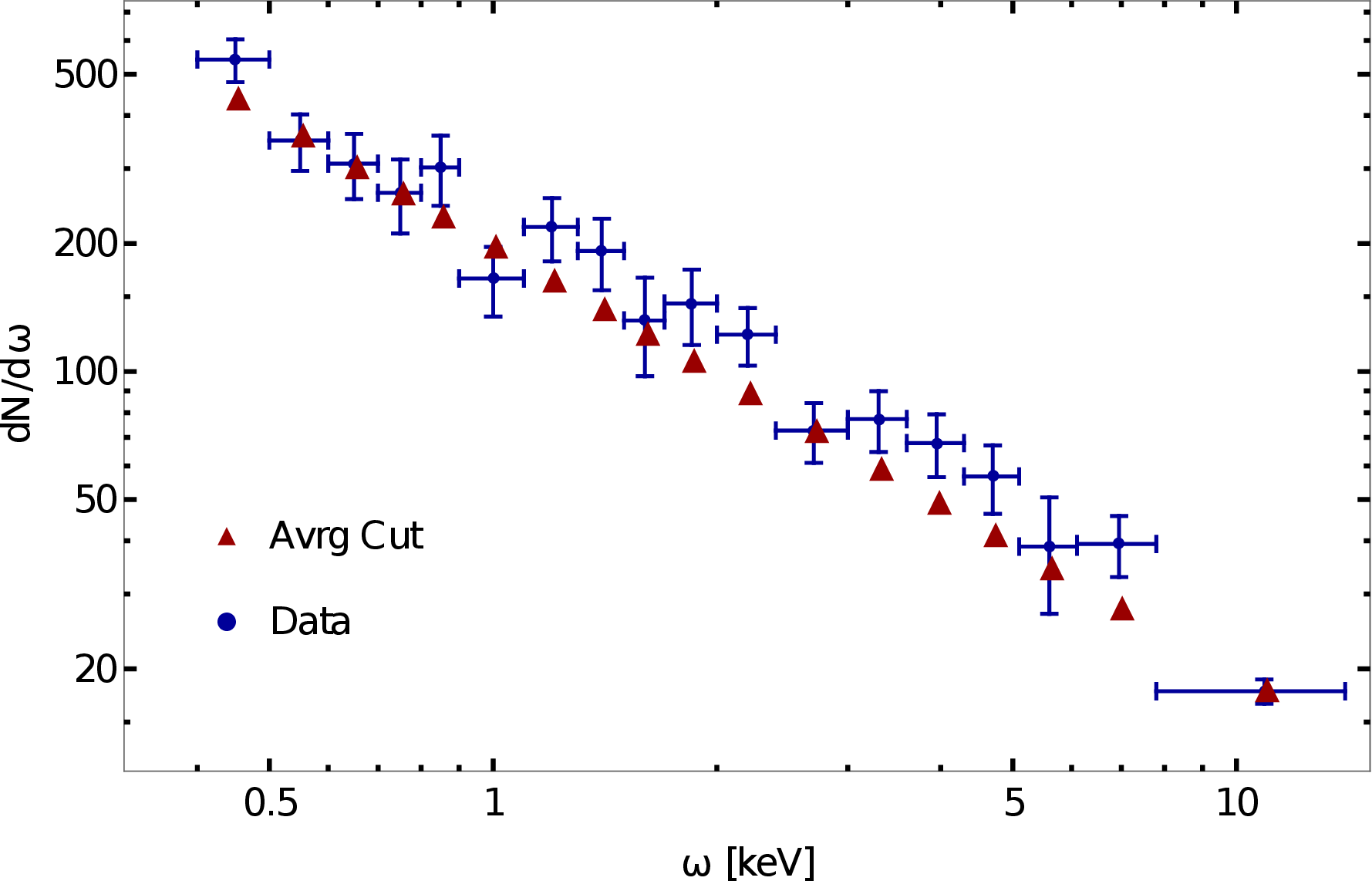}}
	\qquad
	\subfloat[BGO bins]{\label{fig:distr_BGO_bins}\includegraphics[width=0.98\columnwidth]{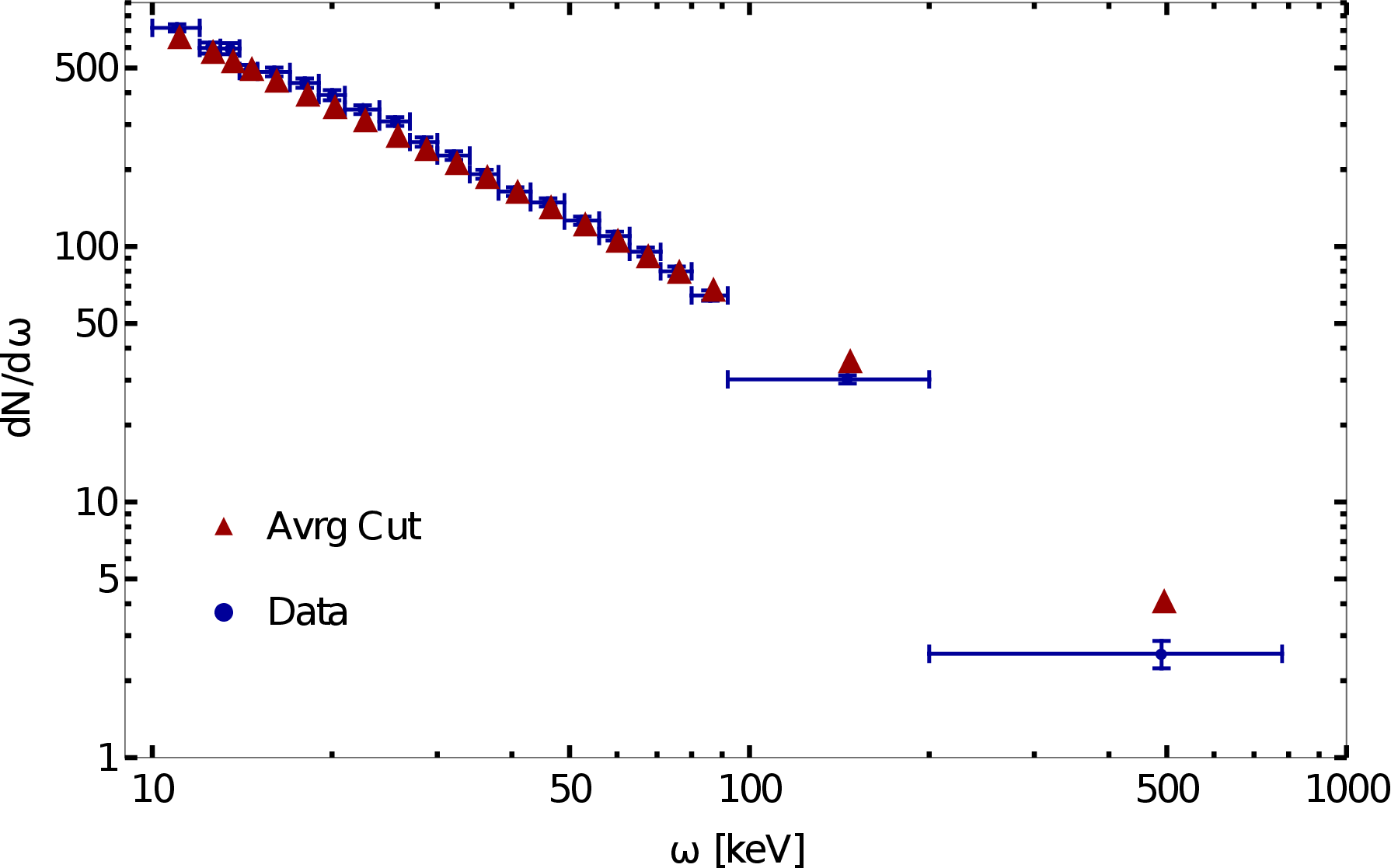}}
\caption{
		Particle distribution for the two detectors.
		The blue dots are the provided experimental data bins.
		The red triangles represent predicted bin heights calculated from the distribution Eq.~\eqref{particle_count_avrg_cut_distr} by using Eq.~\eqref{bin_predict}, without the recoil correction.
	            }
\label{fig:distr_bins}
\end{figure*}

\begin{figure*}[t]
	\subfloat[BGO spectrum with recoil]{\label{fig:distr_BGOrr}\includegraphics[width=0.98\columnwidth]{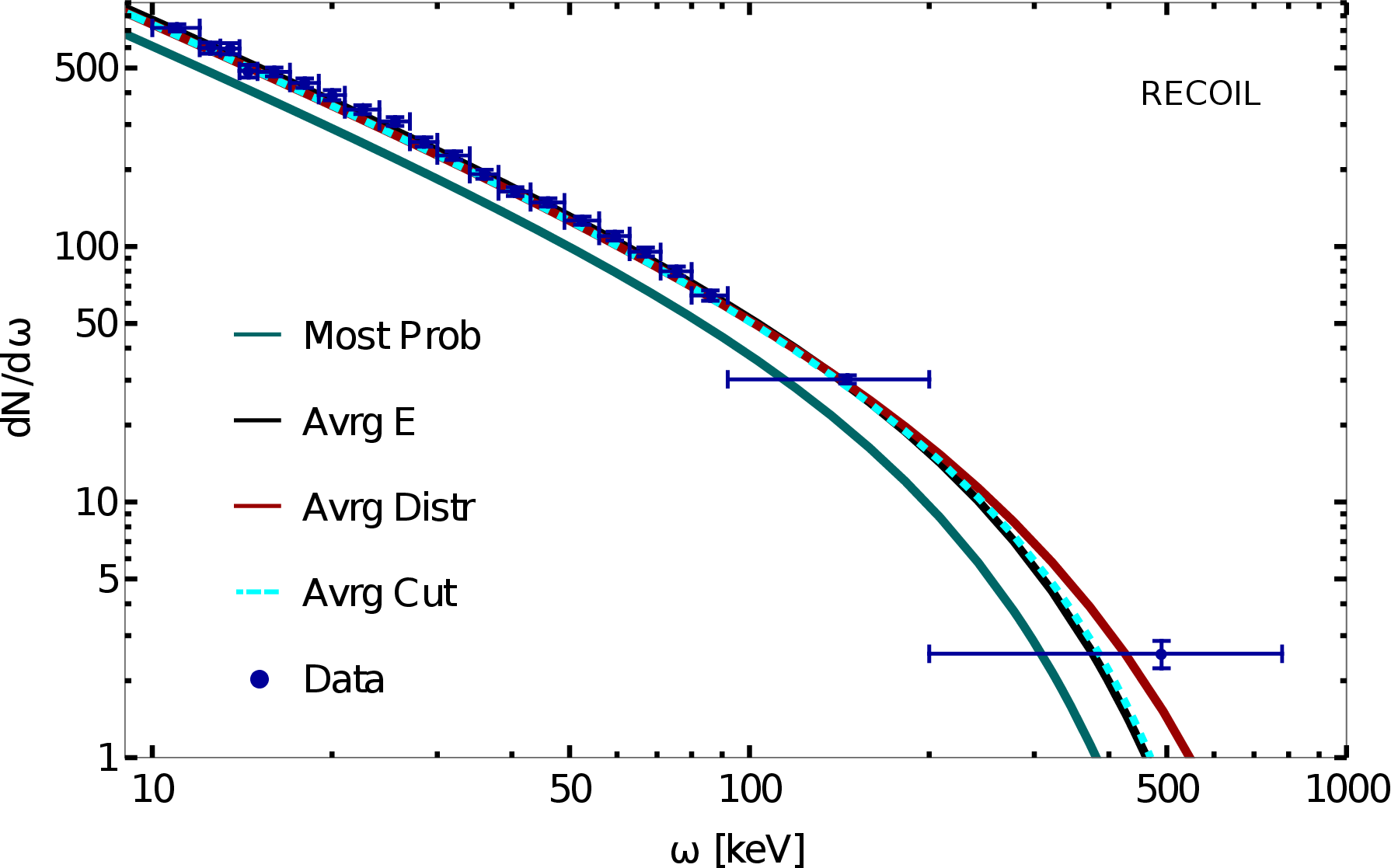}}
	~
	\subfloat[BGO bins with recoil]{\label{fig:distr_BGO_binsrr}\includegraphics[width=0.98\columnwidth]{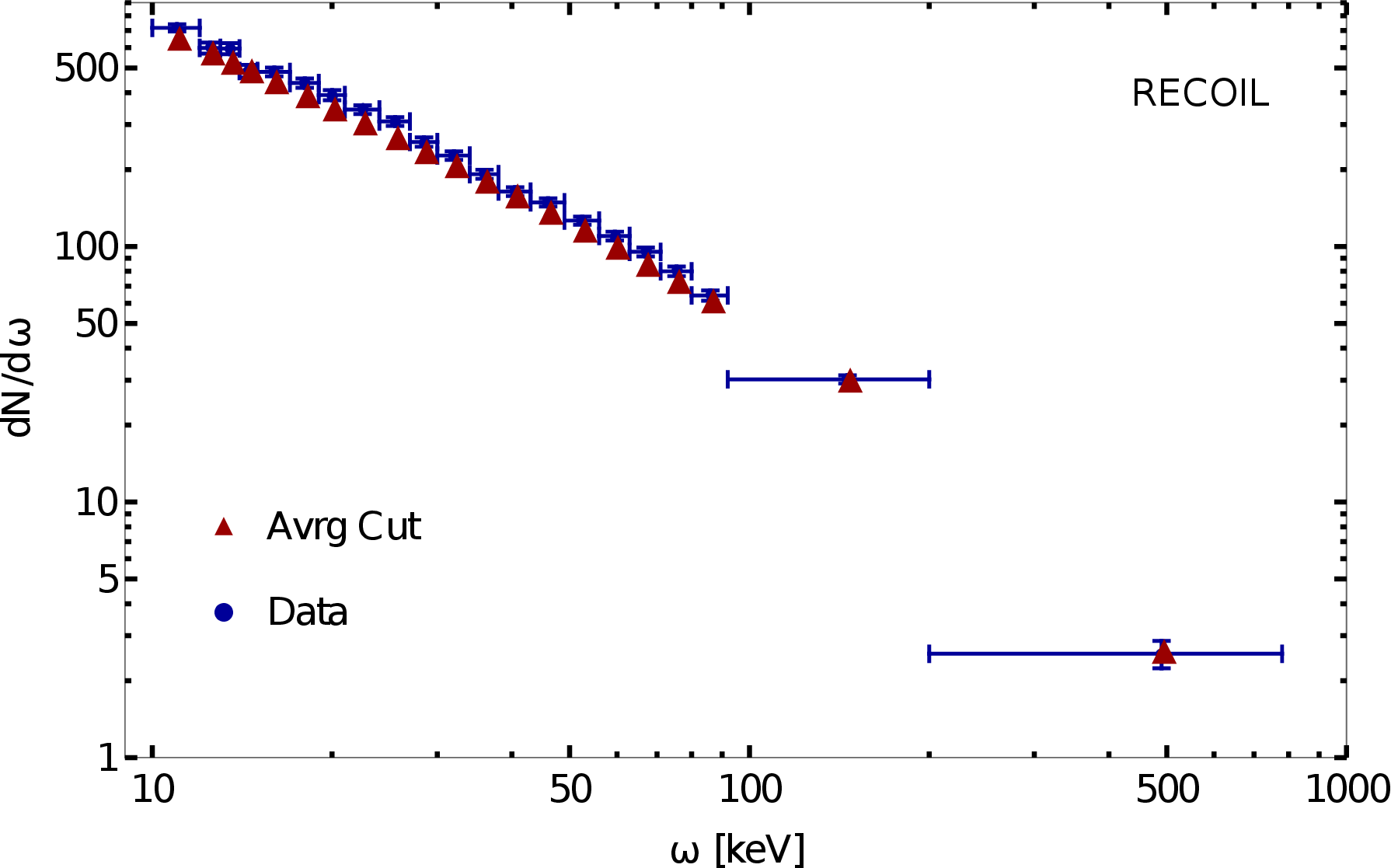}}
\caption{
		Particle spectrum and bin counts for the BGO detector with recoil. The blue dots are the RDK II experimental data.
		The curves are the results of our thermalized 1-D Planckian analysis (without fitting) of Eq.~\eqref{distr_to_compare} and taking into account the chemical potential recoil correction Eq.~\eqref{rr}. All other aspects are the same as in Fig \ref{fig:distr} $\&$ \ref{fig:distr_bins}. Note, the presence of recoil serves to steepen the curve in the high energy tail. 
		\label{fig:distr_rr}
	}
\end{figure*}

The data sets, acquired through the invaluable work of the RDK II Collaboration \cite{RDKII:2016lpd}, encompass the meticulously measured photon spectra.  There they employed two detectors comprised of bismuth germanium oxide (BGO) scintillators and an avalanche photo diode (APD). The low energy spectrum, 0.4 keV to 20 keV, was measured by the APD and the high energy spectrum, 10 keV to 1000 keV, was measured by the BGO. In order to adjust the data to account for the detector response, the data points shown in Figures \ref{fig:distr} and \ref{fig:distr_bins} have been multiplied by the ratio of theory to simulation discussed in \cite{RDKII:2016lpd, pc}. This results in the photon spectrum emitted by the electron during the decay process.

In the following analysis, we will not need to do any fitting; instead, we will directly compute all the parameters of the theoretical curve from first principles. 
We will also use another piece of data --- the branching ratios. The total particle count \cite{RDKII:2016lpd}, along with these branching ratios will then yield our theoretical curves which we compare to the RDK II data.

\subsection{Branching ratios}

Given a photon detector $D$ with detection frequency bandwidth $[\omega_D^\text{low}, \omega_D^\text{high}]$, the branching ratio is defined as a number of electrons that produce a photon in this bandwidth, divided by the total number of electrons from beta decay:
\begin{equation}
	r_D = \frac{N_{\gamma \in D}}{N_{e,\text{tot}}} \,.
\end{equation}
Rephrasing that in probabilistic terms, the branching ratio $r_D$ is the average number of photons inside the bandwidth $[\omega_D^\text{low}, \omega_D^\text{high}]$ produced by one electron.
In terms of a particle distribution $n(\omega)$, we have
\begin{equation}
	r_D = \int\limits_{\omega_D^\text{low}}^{\omega_D^\text{high}} \diff \omega \, n(\omega) \,.
\label{branching_int}
\end{equation}
Using this formula we can compute our theoretical predictions for the branching ratios.
The results are presented in Table~\ref{tab:branch}.

\begin{table*}[t] 
\begin{tabular}{l | l | l | l }
\hline \hline
& APD & BGO & BGO w/recoil\\ \hline
Frequency band, keV & 0.4 -- 14 & 14.1 -- 782 & 
    \\[8pt]
$r_D$, exper. \cite{RDKII:2016lpd} 
	& 0.00582 \shortstack{ \tiny $\pm$ 0.00023 [stat] \\ \tiny $\pm$ 0.00062 [syst]}  
	& 0.00335 \shortstack{ \tiny $\pm$ 0.00005 [stat] \\ \tiny $\pm$ 0.00015 [syst]}
    &
    \\[8pt]
$r_D$, QFT theor. \cite{RDKII:2016lpd,Bernard:2004cm} & 0.00515 (1.0$\sigma$) & 0.00308 (1.7$\sigma$) & 
    \\
$r_D$, our from $n_\text{m.p.}$ Eq.~\eqref{particle_count_mostprob} & 0.00418 (2.5$\sigma$) & 0.00276 (3.7$\sigma$) & 0.00242 (5.8$\sigma$) \\
$r_D$, our from $n_\text{kin-avrg}$ Eq.~\eqref{particle_count_avrg_kin} & 0.00527 (0.8$\sigma$)& 0.00382 (3.0$\sigma$)& 0.00331 (0.23$\sigma$) \\
$r_D$, our from $\bar{n}$ Eq.~\eqref{particle_count_avrg_distr} & 0.00507 (1.1$\sigma$) & 0.00385 (3.1$\sigma$)& 0.00332 (0.22$\sigma$) \\ 
$r_D$, our from $\bar{n}(\omega)_\text{cut}$ Eq.~\eqref{particle_count_avrg_cut_distr} & 0.00507 (1.1$\sigma$) & 0.00361 (1.6$\sigma$)& 0.00322 (0.79$\sigma$) \\ \hline \hline
\end{tabular} \\ 
\caption{Branching ratios. 
Experimental results are from \cite{RDKII:2016lpd}; the standard deviations $\sigma = \sqrt{\sigma_\text{stat}^2 + \sigma_\text{syst}^2}$ for APD and BGO are $0.00066$ and $0.00016$ respectively.
Numbers in parentheses represent deviations from the experimental results in units of $\sigma$.
Our theoretical results are computed with Eq.~\eqref{branching_int};
the third column represents the results computed with the particle distribution Eq.~\eqref{particle_count_general} without the recoil correction, while the forth column uses the recoil-corrected formula Eq.~\eqref{rr}.
The recoil correction plays significant role only at high frequencies and basically has no effect in the APD frequency range.
} 
\label{tab:branch} 
\end{table*}

\subsection{Total electron count and photon distribution}

To compare our particle distribution results with the experiment we will not need to do any fitting to the data.  If we have the total number of electrons $N_{e, \text{tot}}$ produced while gathering the photon data, then the theoretical prediction for the detected photons per keV is given by
\begin{equation}
	\dv{N}{\omega} = N_{e, \text{tot}} \cdot n(\omega) \,,
\label{distr_to_compare}
\end{equation}
where $n(\omega)$ is either the most probable Eq.~\eqref{particle_count_mostprob} or the averaged Eq.~\eqref{particle_count_avrg_distr}.

The RDK II publication \cite{RDKII:2016lpd} presents a comprehensive analysis of the data collection process, highlighting the detection of a substantial number of electron-proton events, amounting to a total of 22 million. However, in order to compare our spectrum to the data, we need an estimate of the total number of measured electrons that made it through all of the detector cuts. We can compute this value, $N_{e, \text{tot}}$, from the available data as follows. 
First, we note that the horizontal error bars $\pm \Delta\omega$ in \cite{RDKII:2016lpd} represent the bin sizes. Then, given a bin width $\Delta_i = 2 \Delta\omega_i$, and a value $(dN/d\omega)_i$ from the data, we calculate the total number of photons in this bin via,
\begin{equation}
	N_{\gamma, i} = (dN/d\omega)_i \ \Delta_i \,.
\end{equation}
Then the total number of photons $N_{\gamma, \text{tot}}$ is given by the sum of $N_i$ over the bins.
The total number of electrons $N_{e, \text{tot}}$ is then calculated using the experimental branching ratio as
\begin{equation}
	N_{e, \text{tot}} = \frac{N_{\gamma, \text{tot}}}{r_D} \,.
\end{equation}
This procedure was carried out separately for each detector.
The resulting counts are presented in Table~\ref{tab:counts}.

\begin{table}[ht] 
\begin{tabular}{l | l | l }
\hline \hline
& APD & BGO \\ \hline
Total photon count & 808 & 17 872 \\
Total accounted electrons & 138 916 & 5 334 982 \\ \hline \hline
\end{tabular} \\ 
\caption{Particle counts. 
} 
\label{tab:counts} 
\end{table} 

Next, using Eq.~\eqref{distr_to_compare} we calculate the particle distribution that we will then compare with the experiment.
We have plotted the resulting curves in Fig.~\ref{fig:distr} $\&$ \ref{fig:distr_rr} for the 1-D Planck spectrum and its recoil correction respectively.

\subsection{Goodness-of-nofit}

Here we perform a statistical test that will help determine how good the no-fit distributions are. One could opt for a reduced $\chi^2$ test, for example. 
However, we should be careful with exactly what it is that is being compared.

Indeed, if we take the theoretical distribution curve, then we would have to compare it with bin-collected data, and some of the bins are pretty wide.
It is somewhat doubtful that the reduced $\chi^2$ test accurately reflects the goodness of prediction in this case.

What we propose to do instead is to use the theoretical curve to construct an object of the same nature as the data, and then compare this object directly to the data. Namely, we make predictions for the bin values.

Given a data bin of frequency span $[\omega_i - \Delta\omega_i, \omega_i + \Delta\omega_i]$, using our theoretical distribution $n(\omega)$ we make a prediction for this bin height as:
\begin{equation}
	n_i^\text{pred} = \frac{1}{ 2 \Delta\omega_i } \int\limits_{ \omega_i - \Delta\omega_i }^{ \omega_i + \Delta\omega_i } \diff\omega \, n(\omega) \,.
\label{bin_predict}
\end{equation}
After that we compare this predicted bin height with the actual bin from the data $n^\text{data}_i$ with uncertainty $\pm \Delta n^\text{data}_i$.
For the plots see Fig.~\ref{fig:distr_bins}.

Thus we are comparing two quantities of the same nature.
This allows us to run a legitimate reduced $\chi^2$ test:
\begin{equation}
	\chi^2 = \frac{1}{\nu} \sum_i \frac{ (n^\text{data}_i - n_i^\text{pred})^2 }{ (\Delta n^\text{data}_i)^2 } \,.
\end{equation}
Here, $\nu$ is the number of degrees of freedom. Since we are not doing any fitting, we take $\nu$ to be the total number of data points (number of bins)%
\footnote{There is some controversy in the subject, see e.g. \cite{andrae2010dos}.}
.
The results for this test are in Table~\ref{tab:test}.

\begin{table}[ht] 
\centering
\begin{tabular}{l | l | l | l}
\hline\hline
& APD & BGO & BGO w/recoil\\ \hline
DOF & \; 18 & \; 21 & \;\;\;\;\;\;\;\;  \\ \hline
$\chi^2$ using $n_\text{m.p.}$ Eq.~\eqref{particle_count_mostprob} 		& \;3.4 & \; 26 & \;\;\;\;\;\;\;\; 37 \\
$\chi^2$ using $n_\text{kin-avrg}$ Eq.~\eqref{particle_count_avrg_kin} 	& \;1.1 & \; 6.5 & \;\;\;\;\;\;\;\; 0.89 \\
$\chi^2$ using $\bar{n}$ Eq.~\eqref{particle_count_avrg_distr} 			& \;1.3 & \; 10 & \;\;\;\;\;\;\;\; 2.8 \\ 
$\chi^2$ using $\bar{n}(\omega)_\text{cut}$ Eq.~\eqref{particle_count_avrg_cut_distr} & \;1.3 & \; 4.2 & \;\;\;\;\;\;\;\; 2.5 \\ \hline\hline
\end{tabular} \\ 
\caption{Reduced $\chi^2$ per degree of freedom.
One can see that the incorporation of recoil yields a better model of the data for the high-frequency portion of the data set. Allow us to emphasize again that there is no fitting of the theory to the data.} 
\label{tab:test} 
\end{table}

\subsection{Comments on the analysis}

Let us make special emphasis on what has been done and what has not been done here.  In particular, with respect to the whole distribution Eq.~\eqref{particle_count_general} which was confirmed without doing any fitting: 
\begin{itemize}
\item 
The fact that we did not fit the $x$-axis confirms the energy-$\kappa$ relation Eq.~\eqref{kappa_from_E}, roughly speaking.

\item
The fact that we did not fit the $y$-axis confirms the distribution prefactor in Eq.~\eqref{particle_count_general}, roughly speaking.

\item Both the distribution data points and the branching ratios were computed from the theory and compared against the data.
\end{itemize}

\paragraph*{Energy conservation:}
The most sophisticated distribution $\bar{n}(\omega)_\text{cut}$ Eq.~\eqref{particle_count_avrg_cut_distr} works best without recoil.
The Heaviside theta function in Eq.~\eqref{particle_count_avrg_cut_distr} serves as a crude manifestation of energy conservation.
In our point-particle semiclassical picture, the electron worldline is a classical trajectory determined by the parameter $\kappa$, which in turn is determined by the maximal kinetic energy of the electron $E_\text{kin}$.
In the event when a photon is radiated, the electron kinetic energy $E_\text{kin}$ is decreased by the amount carried off by the photon $\omega$.
The requirement $\omega < E_\text{kin}$ is the consequence of energy conservation (or, in other words, the requirement that the kinetic energy of the electron cannot turn negative).

The $\bar{n}(\omega)_\text{cut}$ distribution and the \textit{recoil correction} both serve to impose a high-frequency cutoff and therefore account for energy conservation.
Therefore, underlying both methodologies is the conceptually similar incorporation of the conservation of energy based on the kinematics of the system.

\paragraph*{Soft photons:}
In the treatment above, we did not take into account soft photon production.
Photons with frequencies lower than the detection IR cutoff $\omega_\text{IR} = 0.4$ keV are not detected in this experiment.
So, if an electron emits such a photon, it is not registered. However, such an emission will still lower the electron's energy, which in turn affects further photon production.

The effect of soft photon production can be roughly estimated in the following way. If a soft photon with frequency $\omega$ is emitted, it lowers the electron's energy by $\omega$ (recall that in our units $\hbar=1$).
If $n(\omega)$ is the photon count per keV, then the expectation value of the energy carried off by all emitted undetectable soft photons is approximately given by,
\begin{equation}
	\delta E \approx \int\limits_0^{\omega_\text{IR}} \omega \, n(\omega) \, \diff \omega.
\label{soft_correction_1}
\end{equation}
The integrand here is precisely the spectral energy density, $I(\omega) = \omega \,  n(\omega)$. In the IR limit, it approaches the constant
\begin{equation}
	I_\text{IR} = \lim_{ \omega \to 0 } I(\omega) = \frac{e^2}{ 2 \pi^2 } \left(\frac{\eta}{s}-1\right).
\end{equation}
This quantity depends on the electron's energy through $s$, but does not depend on $\kappa$.
This limit is the same for the original density Eq.~\eqref{Iw} and the recoil-corrected density Eq.~\eqref{rr}. Finally, using the detection IR cutoff determined by the APD detector, $\omega_\text{IR} = 0.4$ keV, we have checked that the relative amount of  energy carried off by soft photons is negligibly small,
\begin{equation}
	\frac{\delta E}{E} \approx \frac{ \omega_\text{IR} \, I_\text{IR} }{E} \lesssim 3 \cdot 10^{-6} \,.
\label{soft_correction_2}
\end{equation}
For a plot, see Fig.~\ref{fig:Figure_SoftCorrectionPlot}.
Such a small correction validates the approximation Eq.~\eqref{soft_correction_1}.

\begin{figure}
\includegraphics[width=0.98\columnwidth]{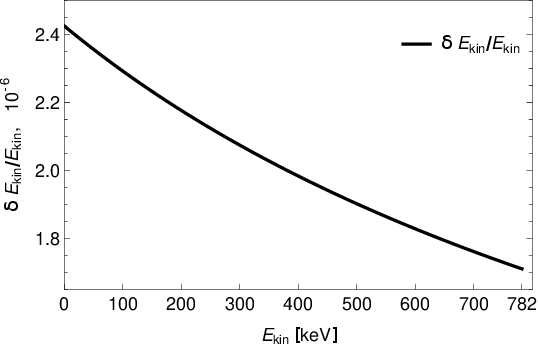}
\caption{
		\label{fig:Figure_SoftCorrectionPlot}
		Correction $\delta E / E$ due to the emission of soft photons, see Eq.~\eqref{soft_correction_1} and Eq.~\eqref{soft_correction_2}.
	 }
\end{figure}
\vspace{10pt}

\paragraph*{Semiclassical picture and QFT:}

The analysis here treats the electron as a classical particle with an exactly and globally defined continuous trajectory.
Using classical electrodynamics, the radiation spectrum is computed;
and after that, semiclassically, electromagnetic waves can be radiated only in discrete packages (photons), which leads to Eq.~\eqref{particle_count_general}.

This analysis does not compete with quantum field theory results, see e.g. \cite{Bernard:2004cm, RDKII:2016lpd, nico}.
Rather, it complements QFT  and can be used to gain different insights. The chi-squared statistic presented by the RDK II collaboration are 1.01 and 1.18 for the APD and BGO detectors, respectively. The fact that these chi-squares are excellent highlights this phenomenon's dual descriptions. In our analysis, we employed techniques related to thermal Larmor radiation. This approach has a few notable features:
\begin{itemize}
    \item It possesses simplicity, both from the computational and conceptual points of view. All the ingredients of the radiative particle distribution --- semiclassical particle distribution Eq.~\eqref{particle_count_general}, recoil correction Eq.~\eqref{rr}, Fermi distribution Eq.~\eqref{decay_Ekin_distr} --- have a clear physical meaning and simple analytic structure.

    \item Because of this simplicity, the one-dimensional Planck factor is experimentally evident, and one can look for observational confirmation (a key finding of this paper).

    \item Another important result is the connection between acceleration and temperature. It is not easy to analyze this connection directly in quantum field theory; the semiclassical picture presented here provides complementarity.

    \item The duality between classical electrodynamics and semiclassical moving mirrors also provides an avenue for investigation of the time dependence of particle creation; see e.g. \cite{good2013time,Good:2022eub}. We leave this for future work.
\end{itemize}

\section{Discussion}

The mathematical equivalence mapping between moving point charge radiation and moving mirror radiation demonstrates a useful and functional connection between radiative beta decay and the dynamical Casimir effect (DCE).  However, the main physical reason for the duality itself is not fully understood.  While an investigation into the underlying symmetry is beyond the scope of this work, it is nevertheless possible to exploit the dual holography (1d $\rightarrow$ 3d) to make predictions about the radiation.

Besides the direct correspondence to the DCE via the duality, the preceding experimental confirmation of classical acceleration temperature may have implications for Hawking radiation. Since the connection between acceleration and temperature can be understood classically, it might be possible to demonstrate that Hawking radiation need not be understood with quantum fields.  Using only a classical formalism, one might appeal to the equivalence principle and investigate Hawking temperature as a relativistic thermodynamic result.

From a more fine-grained perspective, consider that the temperature, $T=1/8\pi M$ \cite{hawking1974black}, is characterized entirely by the black hole mass, i.e. the energy budget of the black hole. When one considers the analog setting of black hole evaporation via the accelerated moving mirror, we find the peel acceleration gives the temperature. Similar to the black hole case, inner bremsstrahlung has a temperature and an acceleration which is also characterized by the maximum energy budget of the system, i.e. the mass difference $E_{\textrm{max}} = \Delta M = m_{n} - m_{p} - m_{e}$ between the neutron and daughter products, $T = \frac{6}{\pi^2}\Delta M$ \cite{Good:2022eub}. So just as the properties of black holes are completely determined by their mass, the kinematics of inner bremsstrahlung is also determined by the mass difference. These similarities highlight the extent to which the moving mirror interpretation of radiative beta decay is an analogy with implications for black hole evaporation.

Another implication for Hawking radiation is that beta decay may act as an experimental black hole analog.  Since the electron emits photons with a 1D Planck spectrum in agreement with the one-dimensional behavior of black holes as predicted by Bekenstein and Mayo \cite{Bekenstein:2001tj}, the data for radiative beta decay could be a valuable source of information about evaporation. The 1D Planck curve is also physically connected to an ordinary resistor through Johnson-Nyquist (white) noise \cite{nyquist}. At low frequencies, this work has confirmed the dimensionality predicted by these systems (for n-dimensional generalizations see \cite{Landsberg_1989}).  Apart from the intrinsic interest of this lower dimensional observation given in the present paper, our findings could also be relevant to the theories of thermal noise in electrical circuits and thermal radiation of the one-dimensional quantum field of the moving mirror model (e.g. Carlitz-Willey \cite{carlitz1987reflections}).

Regardless of these interesting and potential implications, the observational confirmation of the one-channel Planck curve, Eq.~(\ref{Iw}), at the FDU temperature predicted by the dual moving mirror, is the main result of this work.

Interestingly enough, the quantum white noise Planck curve of the resistor signals fundamental thermal radiation and in practice \cite{urick2021fundamentals} is only relevant at high frequencies or low temperatures due to quantum effects.  However, the Planck curve associated here originates in classical electrodynamics (see Sec. IIB of \cite{Ievlev:2023inj}).  Subsequently, as we have emphasized, this work confirms a classical connection between acceleration and temperature. 
The fact that we have arrived at a thermal distribution, and indeed confirmed it from the experimental data, agrees with the notion that the presence of thermality is born out of the kinematics and Lorentz invariance of the system \cite{1998PhRvL..80.3436V}.

The utility of classical electrodynamics in this context may be surprising since beta decay is ultimately a quantum process.  The moving mirror accelerates along a classical trajectory even though the field is quantized. The classical trajectory fits the data because it produces the same photon energy dependence as that of lowest order inner bremsstrahlung (IB) in radiative beta-decay; which in turn is the dominant contribution of IB.  Of course, this IB contribution is famously insensitive to structure-dependent effects, as first shown by Francis Low in 1958 (and now known as the Low theorem \cite{PhysRev.110.974, Unterdorfer:2008zz}.) This insensitivity to structure may, in fact, lay at the heart of the effectiveness of this thermal Larmor/moving mirror formalism.

Summarizing, on the whole, there are several benefits to the perspective given here: the dimensional considerations, the classical vs. quantum interplay; the simplicity and time-dependence of the results; and finally, the straightforward connection between acceleration and temperature.

These results demonstrate a prediction of the duality between the 1+1 dimensional massless scalar field moving mirror effect and
the 3+1 dimensional radiating electron. The accelerated
moving mirror model not only yields the experimentally
confirmed temperature, generating the same energy
emission as that of IB, but has functional predictive
power because it also predicts the 1D spectral distribution. The resulting classical
radiation analysis then provides a no-fit photon spectrum
in agreement with measurement, both with and without
recoil. These results confirm the electron-mirror duality
even though the governing physics behind the duality
is an open question and remains hidden behind the
structureless nature of IB implied by Low’s theorem.
Nevertheless, the classical character of this duality has
demonstrated an unexpected utility regardless of whether
or not, another trajectory description could explain the thermality of the data.

We must also point out that ordinary bremsstrahlung has a rigorous interpretation from the point of view of the Unruh effect \cite{Higuchi:1992td, Landulfo:2019tqj}, namely the emission and absorption of zero energy Rindler photons. In light of these results, it is natural to look ahead and ask if there is also a DCE/thermal Larmor interpretation associated with the Unruh effect as well. Perhaps experiments like those of the RDK II collaboration have the potential to shed more light on some of these aspects of radiation emission.   

With an ever-expanding array of experimental systems which are capable of exploring thermalized particle production, we have found an interesting example in the form of radiative beta decay of the free neutron \cite{RDKII:2016lpd}. The rather peculiar presence of the DCE/thermal Larmor in these particle decays also raises the question of other systems which may provide arenas to explore the tenets of quantum field theory in curved spacetime. As an example, there has been a growing notion of entropic gravity \cite{verlinde2011origin}, modified Newtonian dynamics \cite{milgrom1983modification}, and quantized inertia \cite{mcculloch2007modelling} as a candidate for dark matter arising from an Unruh-type effect on cosmological scales. These, and other interesting systems \cite{DiPiazza:2022wzo, Gregori:2023tun}, including analogs \cite{jacquet2020next}, may find the presence of quantum vacuum fluctuations as an underlying mechanism for unexplained or non-intuitive physics.

\section{Conclusions}
We have examined the photon spectrum emitted during the radiative beta decay of the free neutron measured by the RDK II collaboration. Using the electron-mirror duality, we found an excellent agreement between a thermal one-dimensional Planck spectrum and the photon spectrum. The moving mirror model motivated and predicted the 1D Planck spectrum and radiation temperature of the accelerated point charge. A comparison of the data and theory bears confirmation of thermalization, which is robust enough not to need any fitting parameters. 

In conclusion, this research demonstrates that the rapid acceleration of an electron during neutron decay presents a novel system for investigating the intricate interplay between acceleration and temperature, confirming the existence of thermal radiation emitted by an accelerated electron.  \\

\section*{Acknowledgments}
Funding comes in part from the FY2021-SGP-1-STMM Faculty Development Competitive Research Grant No. 021220FD3951 at Nazarbayev University. M. H. L. was supported by the National Research Foundation of Korea under Grants
No. 2017R1A2A2A05001422 and No. 2020R1A2C2008103.  Appreciation is given to the organizers, speakers, and participants of the QFTCS Workshop: May 23-27, 2022. We thank Thomas R. Gentile and Jeffery S. Nico at the Physics Laboratory at the National Institute of Standards and Technology (Maryland, USA) for helpful advice concerning interpretation and analysis of the RDK II Collaboration data.

\bibliography{2307-023-3D-MorganLynch} 
\end{document}